\documentclass[notitlepage,nofootinbib,preprintnumbers,aps,prd,superscriptaddress]{revtex4-1}
\usepackage{amsmath,amssymb,natbib,bm,color}
\usepackage{appendix}
\usepackage{graphicx}
\usepackage{xcolor}

\begin{document}
\preprint{\hbox{UTWI-36-2023}}

\title{Supermassive Black Hole Binaries in Ultralight Dark Matter}
\author{Benjamin C. Bromley}
\email{bromley@physics.utah.edu}
\affiliation{Department of Physics and Astronomy, University of Utah, Salt Lake City, UT 84112, USA}
\author{Pearl Sandick}
\email{pearl.sandick@utah.edu}
\affiliation{Department of Physics and Astronomy, University of Utah, Salt Lake City, UT 84112, USA}
\author{Barmak Shams Es Haghi}
\email{shams@austin.utexas.edu}
\affiliation{Texas Center for Cosmology and Astroparticle Physics, Weinberg Institute for Theoretical Physics, Department of Physics, The University of Texas at Austin, Austin, TX 78712, USA}

\begin{abstract}
We investigate the evolution of supermassive black hole (SMBH) binaries and the possibility that their merger is facilitated by ultralight dark matter (ULDM). When ULDM is the main dark matter (DM) constituent of a galaxy, its wave nature enables the formation of massive quasiparticles throughout the galactic halo. Here we show that individual encounters between quasiparticles and a SMBH binary can lead to the efficient extraction of energy and angular momentum from the binary. 
The relatively short coherence time of ULDM provides a steady-state population of massive quasiparticles, and consequently a potential solution to the final parsec problem. Furthermore, we demonstrate that, in the presence of stars, ULDM quasiparticles can also act as massive perturbers to enhance the stellar relaxation rate locally, replenish the stellar loss cone efficiently, and consequently resolve the final parsec problem. 
\end{abstract}

\maketitle

\section{Introduction} 
Most, if not all, galaxies in the universe host black holes with masses of  $10^{6}-10^{11}\,M_\odot$~\cite{Richstone:1998ky, Kormendy:2013dxa}.
While the details of the formation and evolution  of these supermassive black holes (SMBHs) and their interactions with the host galaxies are not well understood, a plausible explanation is based on hierarchical galaxy formation. In this scenario, during the merger of smaller galaxies to form the bigger ones, the SMBHs residing in them sink toward the center of the merged galaxy via dynamical friction, mainly driven by weak gravitational interactions with the ambient field of stars. The gravitational settling of SMBHs to form a bound pair can take several Gyr, as their separation diminishes from kiloparsecs to parsecs. After reaching a separation $\sim$ pc, dynamical friction becomes inefficient in ``hardening'' the binary by bringing the two SMBHs  closer together. In this phase, the role of three-body interactions between the binary and individual stars becomes important. A star which approaches the binary at a distance of the order of the binary separation (or smaller), interacts strongly with the binary, and usually is ejected by the slingshot effect at an average velocity comparable to the  orbital velocity of the binary while removing energy from it~\cite{Milosavljevic:2002bn,Yu:2001xp}. Only stars with sufficiently low angular momentum can intersect the binary semi-major axis, and they occupy a small region in the phase space known as the loss cone. Successive encounters of this kind lead to binary hardening at the cost of depleting the loss cone. Under efficient stellar hardening, the binary’s separation can be reduced to $\sim 10^{-2}-10^{-3}$ pc, at which point emission of gravitational waves can take over, resulting in completion of the merger in less than a Hubble time. 

During a merger, the SMBH binary emits nanohertz gravitational waves, which can be probed by pulsar timing arrays (PTAs). The latest data released by PTA collaborations, including the North American Nanohertz Observatory for Gravitational Waves (NANOGrav)~\cite{NANOGrav:2023gor}, the European Pulsar Timing Array (EPTA) along with the Indian Pulsar Timing Array (InPTA)~\cite{EPTA:2023fyk}, the Parkes Pulsar Timing Array (PPTA)~\cite{Reardon:2023gzh}, and the Chinese Pulsar Timing Array (CPTA)~\cite{Xu:2023wog}, indicates evidence for a correlated stochastic gravitational wave signal at nanohertz frequencies. While this signal can be explained by several different models which usually rely on new physics~\cite{NANOGrav:2023hvm}, a cosmic population of SMBH binaries provides a natural source~\cite{NANOGrav:2023hfp}. 
While the new data motivate an explanation based on mergers of SMBH binaries, the occurrence of these mergers on a reasonable time scale may be challenging from the theoretical point of view.  
In the hardening phase, any e-folding of shrinkage in binary separation requires ejecting out a stellar mass of the order of the mass of the binary itself from the central region of the galaxy~\cite{Milosavljevic:2002ht}. A binary needs shrinking by a factor of $\sim 100$ at a separation of $\sim$ 1pc to reach the phase when the emission of gravitational waves dominates, but after a reduction of the order $\sim$ 10 in the size of the binary, the mass of the loss cone drops below the mass of the binary and the binary decay stalls. This is the basis of the final parsec problem~\cite{Milosavljevic:2002bn,Barausse:2020kjy,Milosavljevic:2002ht} which can only be overcome if the empty loss cone is replenished at a sufficiently fast rate.

There have been some proposed mechanisms for field stars to diffuse into the loss cone, such as  Brownian motion of the SMBHs due to discrete
encounters with stars at the center of the galaxies~\cite{Quinlan:1997qe,Yu:2001xp,Milosavljevic:2002bn}, tidal forces in non-spherical (axisymmetric and triaxial) galaxies~\cite{Yu:2001xp,Holley-Bockelmann:2006gbs,Berczik:2006tz,2011ApJ...732...89K,2015ApJ...810..139H,2015ApJ...810...49V,2017MNRAS.464.2301G}, the gas in the central regions of the post-merger galaxy~\cite{Escala:2004jh},  and massive perturbers such as molecular clouds and globular clusters~\cite{Perets:2006bz,Perets:2007nc}.
In consideration of NANOGrav 15-year data set, any new and universal mechanism that facilitates the
merger of SMBH binaries is intriguing and worthy of pursuing. 
In this paper, we examine the effect of dark matter (DM) halos on the formation of SMBH binaries. Specifically, we focus on ultralight dark matter (ULDM) which develops unique features during halo formation that can possibly ameliorate the final parsec problem.  

ULDM is an interesting class of DM models in which the particle DM is a very light boson with a mass in the range $10^{-25}\,{\rm eV} \lesssim m_{\rm ULDM}\lesssim 1\,{\rm eV}$ (the lower and upper bounds on the mass of ULDM can be found in Refs.~\cite{Rindler-Daller:2011afd} and \cite{Berezhiani:2015pia}, respectively).
Subgalactic and small-scale structure prefers a mass range of $10^{-22}-10^{-18}\,{\rm eV}$ for ULDM, which has been shown to be in tension with constraints from the Lyman-$\alpha$ forest~\cite{Armengaud:2017nkf,Irsic:2017yje,Nori:2018pka,Rogers:2020ltq} and the stellar dispersion of ultra-faint
dwarfs~\cite{Marsh:2018zyw,Dalal:2022rmp}. A model-independent lower bound on the mass of DM produced after inflation was found to be $ m_{\rm ULDM}\gtrsim 10^{-18}\,{\rm eV}$~\cite{Amin:2022nlh}.
For recent reviews on ULDM, see~\cite{Hui:2021tkt,Ferreira:2020fam} and and references therein.

The light mass of the ULDM, and consequently its astrophysically large de Broglie wavelength, leads to huge occupation numbers of the ULDM density field. Therefore ULDM is represented as a classical wave (a complex scalar field), described by a wave function that is governed by the Schr\"{o}dinger equation (as the non-relativistic limit of the Klein-Gordon equation) for a particle in a gravitational potential that is sourced by the density of the ULDM field itself via the Poisson equation. The density of the ULDM field is proportional to the square of the modulus of the wave function.

The wave nature of ULDM gives rise to specific, unique features in a DM halo compared to classical point particles. A ULDM halo consists of a central solitonic core of the size of the de Broglie wavelength with an almost constant density that represents the ground state solution to the Schr\"{o}dinger equation. The excited states, on the other hand, add up and surround the core as a Navarro-Frenk-White (NFW) density profile~\cite{Schive:2014dra}.
Therefore, while ULDM mimics the behaviour of cold DM at large scales (larger than the de Broglie wavelength), it leads to a cored profile at smaller scales. The wave interference outside of the core results in the formation of density fluctuations in the NFW-envelope of order unity. The fluctuations that appear as granules in simulations can be described as quasiparticles with an effective mass contained in a de Broglie volume~\cite{Schive:2014dra,Hui:2016ltb} (for a recent study on the validity of this description, see ~\cite{Zupancic:2023qgj}). These quasiparticles are stable as long as the interference pattern is in phase. Therefore, the coherence (the de Broglie) time ($\sim$ time for a boson to cross the length scale of a de Broglie wavelength) of wave patterns determines an effective finite lifetime for the granules~\cite{Veltmaat:2018dfz}. 
Since the coherence time scale can be significantly shorter than the age of the Universe, the quasiparticles developed in the ULDM halo can effectively replenish the loss cone and provide an efficient way for the SMBH binary to lose energy through encounters with them.

The key feature of ULDM that may alleviate the final parsec problem is the interference patterns, i.e., quasiparticles appearing in the halo which have finite lifetimes that can be much shorter than the age of the Universe. The relatively fast coherence timescale may provide an efficient mechanism to replenish the loss cone by generating random fluctuations in the gravitational potential around the binary caused by the formation and disappearance of massive quasiparticles. In this scenario, dynamical friction from ULDM alone may account for formation of SMBH binary. Since dynamical friction from ULDM is suppressed when the size of the system is less than the de Brogle wavelength, formation of the binary needs to take place outside of the solitonic core of the halo. 

As we show in this paper, when the coherence time of ULDM becomes comparable with orbital period of the SMBH binary, individual encounters between quasiparticles and the binary become important. Our simplified numerical analysis shows that these individual encounters can be comparable to individual encounters of the binary with field stars in hardening it, with one critical difference: due to the short coherence time, quasiparticles are effectively replenished, and can result in many encounters of quasiparticles with the binary on the relevant timescale, while field stars need to be brought back into the loss cone, which can take a very long time. 

For a given black hole mass, a lower bound on the ULDM mass comes from the prevalence of quasiparticles and their effect on the dynamical heating of the binary. By increasing the mass of ULDM, eventually the de Broglie wavelength becomes comparable with the Schwarzschild radius of the SMBHs, and therefore ULDM can be treated as cold particle-like DM; this provides an upper bound on the mass of ULDM particles.

While mergers of SMBH binaries purely through interactions with ULDM seems promising, a detailed simulation of the evolution of the SMBH binary near the core, and its impact on both the core and halo is needed. Our study provides the ULDM mass range within which the dynamical friction and individual encounters between quasiparticles and SMBH binary can bring black holes close enough to merge under a Hubble time.      

Although solving the final parsec problem solely with the presence of ULDM seems plausible, 
a more realistic scenario would include the interplay between ULDM quasiparticles and stars, which could provide an efficient way to replenish the stellar loss cone on a reasonable time scale. It is well known that massive perturbers, such as giant molecular clouds or stellar clusters, expedite local relaxation by orders of magnitude relative to two-body stellar relaxation, and consequently replenish the loss cone efficiently by scattering stars into the orbit of the SMBH binary~\cite{Perets:2006bz,Perets:2007nc}. The only requirement for efficient accelerated relaxation by massive perturbers is the existence of a steady state population of large enough inhomogeneities in the galactic mass distribution.
ULDM fluctuations can provide a universal class of massive perturbers in all  galaxies. The universality of this new class of massive perturbers cannot be overemphasized; although giant molecular clouds are common in the disks of spiral galaxies, they do
not survive in elliptical galaxies as a result of a history of major mergers. 
Here, we investigate the mass range of ULDM which can ameliorate mergers of SMBH binaries by replenishing stellar loss cones efficiently.

The outline of this paper is as follows. In \S\ref{sec:bbhevostars} we review the ``classical'' 
theory of black hole binary mergers as driven by interactions with stars. Then, 
in \S\ref{sec:ULDMhalo}, we consider the properties of ULDM galactic halos toward determining the orbital dynamics of black hole binaries within them. Our main results, including novel simulations of binary formation and quasiparticle-binary interactions, are presented in \S\ref{sec:results}. We discuss our findings and conclude in \S\ref{sec:discussandconclude}.

\section{Black Hole Binary Evolution in a Field of Stars}\label{sec:bbhevostars}
Galaxy mergers are expected to lead to the formation of binary black holes. 
The three major physical processes that are involved in the formation and evolution of a binary black hole to reduce the separation between binary components by a few orders of magnitude can be summarized as follows~\cite{Begelman:1980vb}:
\begin{enumerate}
\item 
{\it Binary formation:} during the merger of galaxies, black holes sink toward the center of the new larger galaxy, as a consequence of dynamical friction caused by many small-angle encounters with stars, to form a binary;
\item
{\it Hardening of the binary:} dynamical friction becomes less important and the binary shrinks further through individual large-angle scatterings with stars and gravitational slingshot interactions;
\item 
{\it Gravitational radiation:} when the binary is small enough, emission of gravitational waves becomes relevant and drive the binary towards to rapid coalescence on a reasonable time scale. 
\end{enumerate}
In this section, we review these three different phases of SMBH binary evolution when black holes are surrounded by a classical medium, e.g., stars.
\subsection{Binary Formation: Dynamical Friction from Stars}
If a massive object (star, black hole, or dwarf galaxy, for example) moves within a sea of small, classical particles (stars or DM), its gravity deflects the small particles as they travel past it, creating a wake. The back-reaction is a drag force that slows the massive object's motion through the sea. The contribution to this drag force from a single small particle comes from Rutherford scattering~\cite{2008gady.book.....B};  Chandrasekhar's formula for dynamical friction~\cite{Chandrasekhar:1943ys} accounts for the entire sea. When the mass of the massive object is $M$ and its velocity relative to the sea is $\textbf{v}_M$, while the small particles have mass $m$ and velocity dispersion $\sigma$, this formula is: 
\begin{equation}
  \textbf{F}_{\text{DF}}=-4\pi G^2 M^2 \rho\, \text{ln}\,\Lambda\left[\text{erf}(X)-\frac{2X}{\sqrt{\pi}}e^{-X^2}\right]\frac{\textbf{v}_{M}}{v_M^3},
  \label{eq:ClDF}
\end{equation}
where 
$X=v_M/(\sqrt{2}\sigma)$, $\rho$ is the background  mass density, erf is the error function, and $\text{ln}\Lambda\simeq \text{ln}\left(b_\text{max}/b_{90}\right)$
is the Coulomb logarithm, with 
the impact parameter corresponding to a deflection angle of  $90^{\circ}$ given by $b_{90}=G(M+m)/v_\infty^2$ where $v_\infty$ is the initial relative velocity of the encounter, and the maximum impact parameter given by $b_\text{max}$ which is generally equivalent to the size of the system. Here, we adopt $\Lambda =b_{\rm max} \sigma^2 b_\text{max}/GM$, assuming that the speed of the massive object through the sea is not significantly greater than the particle velocity dispersion, recognizing that the Coulomb logarithm only characterizes the effects of finite-size of the particle sea with respect to the radius of influence of the massive object, but does not rigorously define them (e.g. \cite{2008gady.book.....B}).

For two black holes with masses $M_1$ and $M_2$ traveling through a galaxy, the equations of motion can be written as:
\begin{eqnarray}
    \nonumber M_1\frac{d^2\textbf{r}_1}{dt^2}&=&-GM_1M_\text{galaxy}(r_1)\frac{\textbf{r}_1}{r_1^3}-GM_1M_2\frac{\textbf{r}_1-\textbf{r}_2}{|\textbf{r}_1-\textbf{r}_2|^3}+\textbf{F}_{\text{DF},1},\\
    M_2\frac{d^2\textbf{r}_2}{dt^2}&=&-GM_2M_\text{galaxy}(r_2)\frac{\textbf{r}_2}{r_2^3}-GM_1M_2\frac{\textbf{r}_2-\textbf{r}_1}{|\textbf{r}_1-\textbf{r}_2|^3}+\textbf{F}_{\text{DF},2},
    \label{eq:eqofmotion}
\end{eqnarray}
where $\textbf{r}_1$ and $\textbf{r}_2$ denote the positions of black holes with respect to the center of the galaxy, $M_\text{galaxy}(r_i)$ is the galaxy mass enclosed within a sphere of radius $r_i$, and $\textbf{F}_{\text{DF},i}$ represents the dynamical friction experienced by the $i$th black hole during its motion within the galaxy.

Solving Eq.~(\ref{eq:eqofmotion}) cannot be done analytically in general. But under some circumstances, the evolution of the distance between SMBHs can be followed analytically. Assuming that the density of stars is constant, $\rho(r)=\rho$, and by using Eq.~(\ref{eq:ClDF}), we find the following equation for evolution of the distance between two SMBHs, $\textbf{r}\equiv\textbf{r}_1-\textbf{r}_2$:
\begin{equation}
 \frac{d^2\textbf{r}}{dt^2}
  +\frac{4\pi\rho G^2}{(\sqrt{2}\sigma)^3}\text{ln}\left(\frac{r_i \sigma^2}{GM}\right)\left[M_1f(X_1)\frac{d\textbf{r}_1}{dt}-M_2f(X_2)\frac{d\textbf{r}_2}{dt}\right]
  +G\left(\frac{4\pi}{3}\rho+\frac{M_1+M_2}{r^3}\right)\textbf{r}=0,
  \label{eq:eqdistancecl}
\end{equation}
where $f(X)=\left[\text{erf}(X)-\frac{2X}{\sqrt{\pi}}e^{-X^2}\right]/X^3$, $X_i\equiv v_i/(\sqrt{2}\sigma)$, and $r_i$ is the initial size of the binary.
The function $f(X)$ is a decreasing function of $X$ which satisfies $f(X)\leq f(0)=4/(3\sqrt{\pi})$. Assuming $v_i\ll\sigma$, we replace $f(X)$ with $f(0)$ in Eq.~(\ref{eq:eqdistancecl}). Then for SMBHs with the same mass, i.e., $M_1=M_2=M$, we obtain:
\begin{equation}
   \frac{d^2\textbf{r}}{dt^2}
  + \frac{4\sqrt{2\pi}}{3}\text{ln}\left(\frac{r_i \sigma^2}{GM}\right)\frac{\rho G^2 M}{\sigma^3}\frac{d\textbf{r}}{dt}
  +G\left[\frac{4\pi}{3}\rho+\frac{2M}{r^3}\right]\textbf{r}=0.
  \label{eq:eqdistanceapproxcl}
\end{equation}
The coefficient of the second term in Eq.~(\ref{eq:eqdistanceapproxcl}) reveals the time scale of the evolution of the system, $\tau_{\rm cl}$, given by
\begin{equation}
    \tau_{\rm cl}=\frac{3}{4\sqrt{2\pi}}\frac{1}{\text{ln}\left[r_i \sigma^2/(GM)\right]}\frac{\sigma^3}{\rho G^2 M}.
    \label{eq:taucl}
\end{equation}

To better understand this time scale, we notice that as long as the mass of the galaxy enclosed by the SMBH binary orbit is larger than the mass of SMBHs themselves, i.e., for $r>r_\text{eq}\equiv\left[3M/(2\pi\rho)\right]^{1/3}$, Eq.~(\ref{eq:eqdistanceapproxcl}) can be approximated as:
\begin{equation}
   \frac{d^2\textbf{r}}{dt^2}
 + \frac{1}{\tau_{\rm cl}}
   \frac{d\textbf{r}}{dt}
  +\frac{4\pi}{3}G\rho\textbf{r}=0,  
  \label{eq:3DSHO}
\end{equation}
which is equivalent to the equation of motion for a three-dimensional damped harmonic oscillator with a relaxation time equal to $2\tau_{\rm cl}$ during which $r$ reduces by a factor of $e^{-1}$. The same type of equation has been 
applied to the orbital decay of a single satellite toward the center of its host galaxy due to classical dynamical friction~\cite{2007AmJPh..75..139A}. 

When the size of the system shrinks below $r_\text{eq}$, i.e., for $r<\left[3M/(2\pi\rho)\right]^{1/3}$, the gravitational force between SMBHs dominate the gravitational force due to stars, and consequently Eq.~(\ref{eq:eqdistanceapproxcl}) can be approximated as:
\begin{equation}
   \frac{d^2\textbf{r}}{dt^2}
  + \frac{1}{\tau_{\rm cl}}
   \frac{d\textbf{r}}{dt}
  +\frac{2GM}{r^3}\textbf{r}=0,
  \label{eq:decayorbit}
\end{equation}
which simply describes the orbital decay of a two-body system in the presence of friction. 
In this phase of binary evolution, 
the black holes each independently interact with the sea of stars. Once the orbits decay to the point where $v_M \gtrsim$$\sigma$, then close encounters --- not interactions with the swarm of stars --- between the SMBH binary and the stars become
important, as we describe next.
\subsection{Hardening of the Binary}
\label{subsec:hardstar}
When the mass enclosed by the orbit of the SMBH binary is comparable
to the mass of the binary itself, the hardening phase of the binary begins.  
At this stage, the binding energy (or equivalently, the kinetic energy) per unit mass of the binary approaches the kinetic energy per unit mass of the stars in the galaxy, i.e., $\sigma^2$. 
Therefore the semi-major axis of a binary at the beginning of the hardening phase can be defined as:
\begin{equation}
    a_h\equiv \frac{G\mu}{4\sigma^2},
\end{equation}
where $\mu=M_1M_2/(M_1+M_2)$ is the reduced mass of the binary.
Once the separation between the black holes drops below $a_h$, losing energy by dynamical friction becomes inefficient. Stars that encounter the binary at a distance $\sim a \lesssim a_h$, are expected to undergo a gravitational slingshot and be ejected at a velocity $v\sim \sqrt{G(M_1+M_2)/a}$, which is the typical orbital velocity of the
binary~\cite{2015ApJ...810...49V}.

Each individual encounter between stars of mass $M_\star$, and the bound binary with binding energy $E=GM_1M_2/2a$, leads to a fractional energy change of $\delta E/E\sim \xi M_\star/(M_1+M_2)$, where $0.2\lesssim\xi \lesssim 1$~\cite{1980AJ.....85.1281H, Quinlan:1996vp}. A noticeable change in the energy of the binary requires a large number of stars (of the order of $(M_1+M_2)/M_\star$) encountering the binary~\cite{2014SSRv..183..189C}. 
The cross-section for three-body encounters between the binary and stars can be defined as $\Sigma\equiv 2\pi(M_1+M_2)Ga/\sigma$~\cite{Celoria:2018mzr}. And therefore, by assuming a fixed background of stars with density $\rho_\star$, one obtains the hardening rate,
\begin{equation}
    \frac{d}{dt}a^{-1}\sim \xi\pi\frac{G\rho_\star}{\sigma},
\end{equation}
and consequently, the  hardening time scale,
\begin{equation}
    \tau_{\rm hard} \sim \frac{\sigma}{\pi G\rho_\star a}.
\end{equation}
Since the hardening time scale increases while the binary is shrinking, even for a fixed background stellar field, the binary merger could stall. Worse still, the population of stars with orbits that get close to the binary are scattered away; these stars, and the scattering processes that remove them from their original phase-space positions, define the loss cone. In the case of hardening with a constant rate, an overall stellar mass comparable to the mass of the binary is removed from the loss cone, and is no longer available to play a role in hardening the binary~\cite{2013CQGra..30x4005M}. 

Binary hardening by stellar encounters thus depends on mechanisms that replenish the loss cone. The diffusion of star by mutual scattering is one possibility. As a measure of the efficiency of this process, we consider the two-body relaxation time during which a star's velocity diffuses by an amount comparable to the velocity itself. The duration of each crossing at a distance $r$ from the center of the galaxy is given by $t_\text{cross}=r/\sigma$, which leads to following relaxation time scale in a galaxy with $N$ total stars~\cite{2008gady.book.....B}:
\begin{equation}
    t_\text{relax}\simeq \frac{0.1 N}{\text{ln}\,N} t_\text{cross}.
    \label{eq:classrelax}
\end{equation}
A more accurate relaxation time can be estimated from the Fokker-Planck equation in terms of the mass of the star, $M_\star$, and the local density of stars, $\rho_\star(r)$, as~\cite{2008gady.book.....B}:
\begin{equation}
    t_\text{relax}\simeq 0.34\frac{\sigma^3}{G^2 M_\star\rho_\star(r) {\rm ln}\Lambda},
    \label{eq:classrelax2}    
\end{equation}
where $r$ is the size of the binary and 
$b \equiv b_\text{max}$ 
is the larger value of $GM_\star/\sigma^2$ and the size of the stars. It turns out that
the relaxation time is much longer than the Hubble time and therefore new dynamical processes are needed to refill the loss cone efficiently.  In the absence of such a mechanism, the merger stalls.
\subsection{Gravitational Radiation and Merging}
Eventually, when the the binary’s semi-major axis is small enough, the emission of gravitational waves 
causes the binary to inspiral and eventually merge. The rate of inspiral 
due to gravitational waves is approximately $d a/dt\sim -a^{-3}$; the time scale for merger via emission of gravitational waves with initial separation distance $\sim a$ is
\begin{equation}
 t_{\rm gr}=\frac{5}{256F(e)}\frac{c^5}{G^3}\frac{a^4}{M_1M_2(M_1+M_2)},
\end{equation}
where $c$ is the speed of light, $e$ is the eccentricity and $F(e)=(1-e^2)^{-7/2}\left(1+\frac{73}{24}e^2+\frac{37}{96}e^4\right)$~\cite{Peters:1964zz}.
For an equal-mass binary, merger in a Hubble time requires $a \lesssim 0.05 a_h$~\cite{Merritt:2004gc}. 

\section{SMBH Binary formation and evolution in a ULDM halo}
\label{sec:ULDMhalo}

We now explore how the quantum-mechanical properties of ULDM, as the main DM constituent of a galaxy, would impact the formation and evolution of a SMBH binary. We begin by reviewing the formation of a ULDM halo and its unique features due to the wave nature of ULDM.

The dynamics of ULDM, as a scalar field of mass $m_{\rm ULDM}$, can be described by the non-relativistic limit of the governing Klein-Gordon equation.  This leads to a classical field description for the ULDM best described by the Schr\"{o}dinger equation for a complex scalar field, $\psi$~\cite{Ruffini:1969qy}:
\begin{equation}
    i\hbar\partial_t\psi=\left[-\frac{\hbar^2 \nabla^2}{2m_{\rm ULDM}}+m_{\rm ULDM} (U_\psi+U_\text{sat})\right]\psi,
    \label{eq:schrodinger}
\end{equation}
where $U_\psi$ is the gravitational potential of the ULDM (self-gravity) obtained from the Poisson equation,
\begin{equation}
   \nabla^2 U_\psi=4\pi G m_{\rm ULDM}|\psi|^2, 
\end{equation}
and $U_\text{sat}$ is the gravitational potential of massive objects (satellites) moving though the ULDM sea. By including self-gravity, it is expected that~\cite{Hui:2016ltb} ULDM halos develop a central soliton as the ground state of the Schr\"{o}dinger equation, which is also the densest state of the system. Numerical studies of the evolution of ULDM halos show that the radius of the soliton core is of the order of the de Broglie wavelength of the velocity dispersion ($\sigma$) of ULDM with mass $m_{\rm ULDM}$, i.e., $\lambdabar_\sigma=\hbar/(m_{\rm ULDM} \sigma)$~\cite{Schive:2014hza}.  

The wave nature of ULDM gives rise to interference between different modes within the halo. Since the density of ULDM in the halo is proportional to the squared modulus of the wave function, the interference fringes can be interpreted as quasiparticles with an effective mass contained in a de Broglie volume, given by~\cite{Hui:2016ltb}:
\begin{equation}
 m_\text{eff} \sim \rho\lambdabar_\sigma^3,
 \label{eq:effmass}
\end{equation}
where $\rho$ is the local density of the ULDM halo. These quasiparticles can also be understood as density fluctuations in the halo which are correlated over a length scale of the order of the de Broglie wavelength. The size of the soliton core is also of the order of the de Broglie wavelength. The interference between different modes within the ULDM halo, which is responsible for appearance and disappearance of quasiparticles, introduces a new time scale, the coherence (de Broglie) time that characterizes the interval during which the interference is in phase. The coherence time scale can be estimated by using the fact that $\lambdabar_\sigma\sim 2\pi/k$ where $k$ represents a typical momentum mode. This leads to a difference in energy of different modes which is of the order of $\delta E\sim k^2/(2m_{\rm ULDM})\sim k\sigma/2$ and can be converted into a time scale~\cite{Hui:2021tkt}. The coherence time scale can be defined as~\cite{Hui:2021tkt}
\begin{equation}
    \tau_c\equiv 2\pi\frac{\lambdabar_\sigma}{\sigma}=\frac{2\pi\hbar}{m_{\rm ULDM}\sigma^2}.
\end{equation}
Note that the prefactor $2\pi$ is arbitrary and sometimes is omitted (e.g., in Ref.~\cite{Veltmaat:2018dfz}).

The temporal correlation function inside and outside of the core, evaluated in~\cite{Veltmaat:2018dfz}, shows that beyond the solitonic core (a distance of $\sim 3.5 r_c$ where $r_c$ is radius of the core), temporal correlation drops. This is a manifestation of the enhanced coherence/stability of the core relative to the quasiparticles in the halo around it; roughly speaking, one can assign a finite lifetime of order of coherence time scale to the quasiparticles~\cite{Veltmaat:2018dfz}. 

In this context, we explore the mechanisms by which SMBHs can lose energy to the ULDM halo, allowing them to form a binary and eventually merge. More specifically, we focus on dynamical friction from ULDM and direct encounters of ULDM quasiparticles with the SMBH binary.  Finally, we explore conditions under which SMBH binaries in ULDM potentially can merge in less than a Hubble time. 

\subsection{Dynamical Friction in a ULDM Halo}
\label{subsec:DFULDM}
For a point-like satellite of mass $M_\text{sat}$ in a ULDM halo with density $\rho$ and mass $m$ per particle, moving with velocity $\textbf{v}=v \hat{z}$ in the halo rest frame, there is an exact time-independent solution to Eq.~(\ref{eq:schrodinger}), which is basically the scattering by a Coulomb potential, given by~\cite{Hui:2016ltb}:
\begin{equation}
    \psi(\textbf{r})=\sqrt{\rho}e^{\pi\beta/2+ i z/\lambdabar}|\Gamma(1-i\beta)|M\left(i\beta,1,\frac{i}{\lambdabar}(r-z)\right),
\end{equation}
where $\beta=G M_\text{sat}/(v^2\lambdabar)$, $\lambdabar=\hbar/(m_{\rm ULDM}v)$ is  the de Broglie wavelength associated with the relative velocity of the satellite and the ULDM medium, and $M(a,b,x)$ is the Kummer function or confluent hypergeometric function of the first kind.  The wavefunction, or equivalently the density of the ULDM, can be used to calculate the dynamical friction (the gravitational force of the point-like satellite on the sea of ULDM) by integrating over a spherical region surrounding the satellite~\cite{Hui:2016ltb}.

In reality, rather than a single plane wave, a wave packet corresponding to a distribution of velocities should be ascribed to the ULDM medium~\cite{ Bar-Or:2018pxz, Lancaster:2019mde}. Although an exact evaluation of the dynamical friction sourced by ULDM demands numerical solutions, for some limiting cases one can obtain analytical expressions as leading terms of expansions of the drag force. The parameter of interest in the expansion is the de Broglie wavelength of ULDM.
For small enough ULDM mass corresponding to a de Broglie wavelength greater than the size of the binary system, $\lambdabar_\sigma\gg b$, the ULDM background can be treated as a constant density, while for 
high ULDM mass where $\lambdabar_\sigma\ll b$, 
we approach the classical limit of ULDM. 
Dynamical friction, for these two extreme regimes, can be expressed as~\cite{Hui:2016ltb, Bar-Or:2018pxz, Lancaster:2019mde}: 
\begin{equation}
\textbf{F}_\text{DF}= - 4\pi\rho G^2M^2\frac{\textbf{v}_\text{rel}}{v_\text{rel}^3}\left\{
        \begin{array}{ll}
            \text{Cin}\left(2b/\lambdabar_\sigma\right)+\frac{\text{sin}\left(2b/\lambdabar_\sigma\right)}{2b/\lambdabar_\sigma}-1 & b\ll \lambdabar_\sigma, \\
             \text{ln}\left[b/(\lambdabar_\sigma/2)\right]\left[\text{erf}\left(\frac{v_\text{rel}}{\sqrt{2}\sigma}\right)-\sqrt{\frac{2}{\pi}}\frac{v_\text{rel}}{\sigma}e^{-\frac{1}{2}\left(\frac{v_\text{rel}}{\sigma}\right)^2}\right] & b\gg \lambdabar_\sigma,
        \end{array}
    \right.
    \label{eq:QDF}
\end{equation}
where $\text{Cin}(x)=\int_0^x dt(1-\text{cos}\, t)/t$ is the cosine integral.

When the size of the system is smaller than the de Broglie scale, the wave nature of ULDM can suppress dynamical friction by smoothing the tail of the particle phase-space distribution in the wake formed behind the massive object passing through the sea of ULDM.
For the choice of parameters that are consistent with the wave limit of ULDM ($b\ll \lambdabar_\sigma$), dynamical friction decreases significantly. This suppression of dynamical friction by ULDM with low mass ($m\sim10^{-22}\,\text{eV}$) has been used to explain the survival of globular clusters against sinking to the center of Fornax dwarf spheroidal galaxy~\cite{Hui:2016ltb, Lancaster:2019mde}.
Comparing Eq.~(\ref{eq:QDF}) with Eq.~(\ref{eq:ClDF}) shows that when the size of the system is much much larger than the de Broglie wavelength of the ULDM ($b\gg \lambdabar_\sigma$), dynamical friction sourced by ULDM follows the form of the classical result, but with a difference; the wave nature of ULDM introduces a softening scale of the order of the de Broglie wavelength~\cite{ Bar-Or:2018pxz}.

When the separation distance between black holes is larger than the ULDM de Broglie wavelength, a time scale similar to Eq.~(\ref{eq:taucl}) can be found for evolution of a binary of equal mass black holes in ULDM due to dynamical friction as:
\begin{equation}
    \tau=\frac{3}{4\sqrt{2\pi}}\frac{1}{\text{ln}\left[r_i/(\lambdabar_\sigma/2)\right]}\frac{\sigma^3}{\rho G^2 M}.
    \label{eq:tauULDM}
\end{equation}

\subsection{Direct ULDM Quasiparticle Encounters with a SMBH binary}
\label{subsec:ULDMSMBHencounter}

As the orbital separation of a SMBH binary shrinks over time, it may approach the de Broglie wavelength, $\lambdabar_\sigma$ of ULDM particles. Then, the approximation of dynamical friction from Eqs.~(\ref{eq:QDF}) and (\ref{eq:eqofmotion}), with each binary partner independently plowing through an unperturbed sea of ULDM, breaks down. We thus consider situations where the binary partners jointly stir the particle sea. We focus on fluctuations of ULDM that yield massive quasiparticles which interact directly with the binary. 
Our goal is to determine how ULDM quasiparticels, in the absence of stars, impact the hardening phase of a SMBH binary and the final parsec problem. Specifically, we are interested in individual encounters between ULDM quasiparticles and the SMBH binary, anologous to three-body inteactions between the binary and stars. This problem is complicated, requiring a separate analysis. Here, we solve a simplified scenario to explore the relevance of this mechanism to the orbital evolution of an SMBH binary.

In our simplified analysis, we represent a ULDM quasiparticle as a wavefunction $\psi$ that is spatially compact, with a spherically symmetric Gaussian waveform (standard deviation parameter, $R$), that modulates the amplitude of an otherwise plane-wave solution with de Broglie wavelength $\lambda = R$.  The packet is initially propagating toward the binary with speed $v_0 = \hbar/m_{\rm ULDM}/\lambdabar$, where $m_{\rm ULDM}$ is the mass of an individual ULDM particle. The binary itself consists of two equal-mass point particles ($M_1 = M_2$), initially on a circular orbit with semimajor axis $a_0$. The wavefunction evolves according to the time-dependent Schr\"odinger equation with the potential set by the gravity of the binary. Each black hole experiences a Newtonian gravitational force from its partner and from the wave packet, whose mass density is $m_\text{eff}|\psi|^2$, where $m_\text{eff}$ is the total mass of the quasiparticle. We ignore the self-gravity of the wave packet.

To illustrate encounters between a quasiparticle and an SMBH binary, we adopt black hole masses $M_1 = M_2 = 5\times 10^6$~M$_\odot$, and a circular binary orbit with initial orbital separation of $a_0 = 1$~pc and randomly chosen phase and orientation.  The quasiparticle's total mass, $m_\text{eff}$ is 1\%\ of the total binary mass, and the mass of the individual ULDM particles is set to $m \approx 3\times 10^{-19}$~eV. The de Broglie wavelength of the particles and the spatial extent of the initial quasiparticle are $2\pi\lambdabar = R = 0.2$~pc. The quasiparticle is aimed at the binary center of mass with a packet speed of $v_0 \approx 100$~km/s when outside of the binary's gravitational influence. With these initial conditions, we solve for the evolution of the wave packet and the binary together. The Appendix provides the details of our calculations.

\begin{figure}
    \centering
    \mbox{\includegraphics[width=6in]{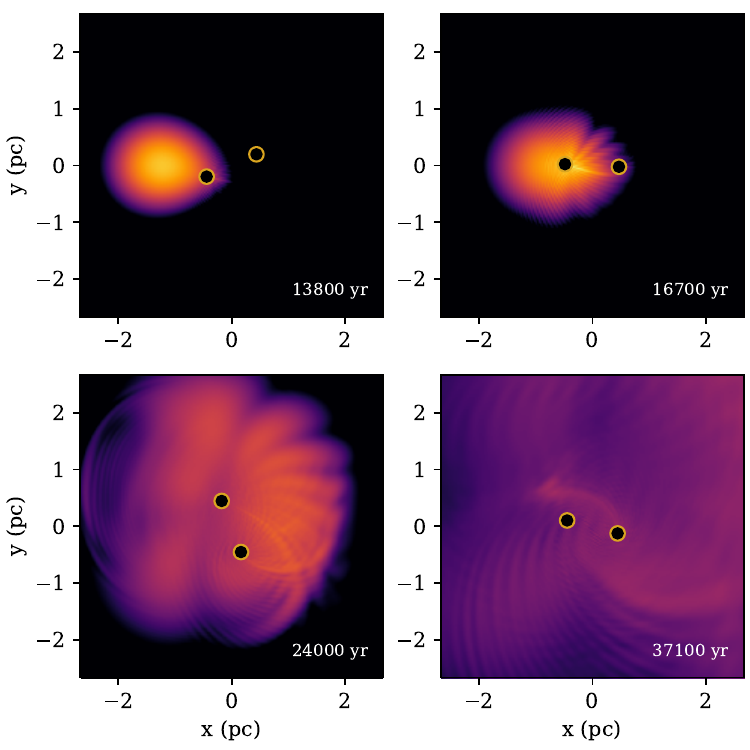}}
    \caption{Snapshots of a dark quasiparticle scattering with a binary black hole. The black holes each have a mass of $5\times 10^6$~M$_\odot$.
    The quasiparticle, with a total mass of 1\%\ of the combined black hole mass, starts at time $t=0$ just to the left of the region shown in the snapshots, progressing in increasing time from left to right, then top to bottom. The color map shows the logarithm of $|\psi|$, 
    down to $10^{-4}$ times its initial peak. Thus the initial wavefunction is more campact than might appear visually in the upper left plot, while in the lower right plot, little of the quasiparticle remains in the vicinity of the black holes. The nuances of the quasiparticle-binary interaction can be seen in this animation: 
    \texttt{www.astro.utah.edu/\~{}bromley/fuzzy\_{}6837802.gif}}
    \label{fig:fuzzysim}
\end{figure}

Fig.~\ref{fig:fuzzysim} illustrates our results for one of a suite of 200 simulations, each with a unique binary orbital phase and orientation. In the Figure, the wavefunction scatters off of one binary partner only to diffract off of the other black hole as it swings around in its orbit. Much of the wavefunction is lost to regions outside the computational domain after this first encounter; the rest lingers near the binary and will eventually be ejected as well. Presumably, some small fraction of the quasiparticle becomes bound by --- and accreted onto --- the black holes.

\begin{figure}
    \centering
    \mbox{\includegraphics[width=6in]{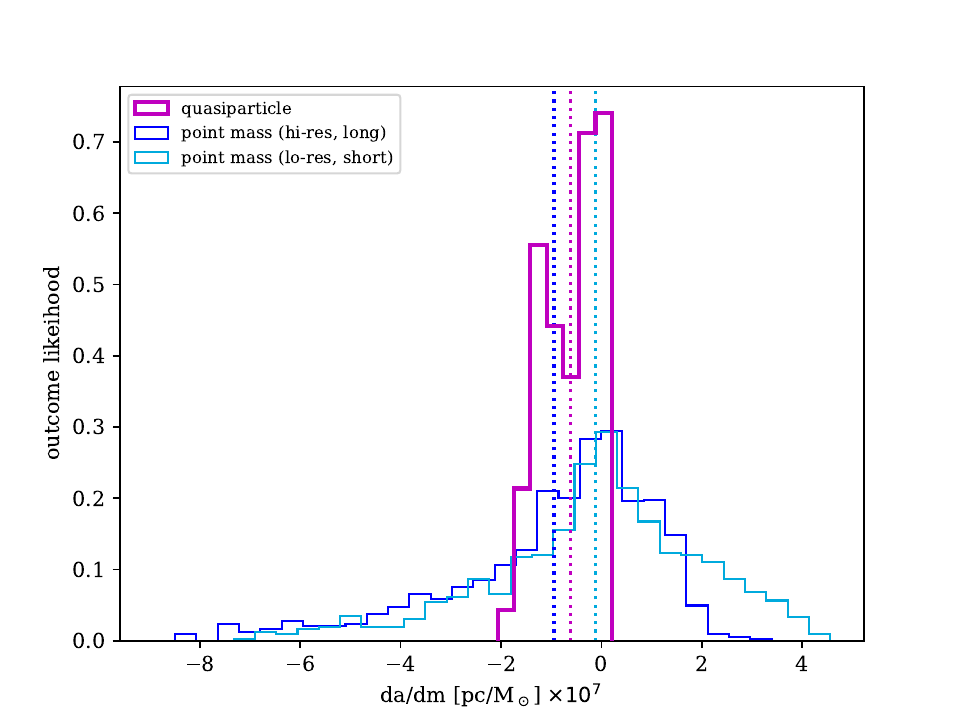}}
    \caption{Histogram showing the change in binary semimajor axis for encounters with ULDM quasiparticles and classical point particles. The change in semimajor is given per encounter with a unit mass scatterer, either a ULDM quasiparticle (magenta histogram) or a point particle (cyan and blue histograms). The simulation times and force smoothing scales are identical for the ULDM quasiparticles and the point particles represented by the cyan histogram. The vertical lines show median values. Comparing these cases, we find that ULDM is more effective per unit (quasiparticle) mass than point particles in driving binary inspiral. If we integrate the point particles for longer time and with less force smoothing (blue histogram), then the effect of point particles on binary inspiral is closer to  that of ULDM.}
    \label{fig:dasemidm}
\end{figure}

Throughout each simulation, we track the change in semimajor axis of the binary in response to the quasiparticle's gravitational influence. Figure~\ref{fig:dasemidm} summarizes the results: at 1~pc separation, an equal-mass binary (total mass $10^7\ M_\odot$) loses on average about $da/dm \sim 10^{-7}$~pc/$M_\odot$ per encounter with a ULDM quasiparticle. We compare this rate of loss of orbital energy to that caused by stars on orbits that are similar to the quasiparticle. For this purpose we use the \texttt{Python} function \texttt{solve\_{}ivp} in the \texttt{SciPy.integrate} module. We ran a thousand trials of stars launched to loosely sample the spread in the quasiparticle wave packet in position and speed. Using the same force smoothing scale and duration as for the quasiparticle simulations, we obtain a much broader spread in outcomes of $da/dm$, with the mean inspiral rate that is a few times slower than with quasiparticles. Reducing the force smoothing and increasing the simulated time of the encounters enhances binary inspiral. The mean binary inspiral from quasiparticle scattering is closer to these ``high-resolution'' results. 

The spread in binary inspiral rates, $da/dm$, from the quasiparticle simulations is much less than from individual stars. This outcome can be roughly understood from the perspective that a quantum mechanical orbit effectively samples many individual stellar orbits simultaneously. Then, the inspiral from a single quasiparticle interaction will be closer to the average $da/dm$ from many single-star encounters. 

The quasiparticle-binary simulations presented here demonstrate that ULDM may well push SMBH binaries inward from parsec-scale orbital separation. However, there are several important caveats to this analysis: First, in this preliminary work, we explore a single type of encounter between a quasiparticle and an equal-mass binary, varying only the orientation and phase of the binary. A more comprehensive analysis to test the robustness of our results would cover a much broader parameter space that includes binary masses and separation, as well as quasiparticle properties ($m$, $R$, $v_0$, starting location relative to the binary, etc.). 

The second caveat is that our numerical method is limited in its spatial and dynamical resolution. Our numerical experiments with variations of grid sizes and force smoothing indicate that we have robustly simulated binary inspiral ($da/dm<0$), however details matter to the exact outcome. Force smoothing in particular generates a comparatively shallow potential well that can affect quantum mechanical scattering on the scale of the computational grid. Future work with more sophisticated algorithms (multigrid/adaptive mesh refinement or specialized coordinates, for example) would better resolve the quasiparticle-binary interactions.

A final word of caution is that we have taken advantage of a scaling relations between the mass of the scatterer, the number of scattering encounters, and the binary inspiral rate by using an unrealistically large quasiparticle mass (10\%\ of the binary mass), and dividing the inspiral rate by that amount, e.g., Fig.~\ref{fig:dasemidm}. This choice balances the need to be in a low-mass regime where the binary inspiral rate per encounter is strictly proportional to the scatterer's mass, along with numerical limitations of measuring orbital changes from small scattering events. By varying scatterer masses (quasiparticles and point particles), we have confirmed that our choice does not negatively impact our results. 

Despite limitations, the calculations presented here reveal the potential importance of ULDM scattering with a black hole binary when the de Broglie wavelength is comparable to, or modestly less than, the binary separation. At least, our results are a hopeful call for more detailed numerical studies.  Meanwhile, with the main result that ULDM, similar to stars, can drive binary inspiral, we next consider the main advantage of ULDM over a stellar population toward the merger of a SMBH binary.

\subsection{SMBH Binary Mergers in ULDM: Challenges and Opportunities}
\label{subsection:obpo}
The classical limit of dynamical friction from ULDM, i.e., when $b\gg\lambdabar_\sigma$, inherits the wave behaviour of ULDM mildly; as one can see from Eq.~(\ref{eq:QDF}), the Coulomb logarithm includes de Broglie wavelength as the softening scale. This simple fact tells us that dynamical friction via ULDM by itself, may be able to reduce the size of the SMBH binary after hardening. A successful merger needs a reduction of the size of the binary by a factor of $\sim 100$ in less than a Hubble time. To this end, the size of the system should stay larger than the de Broglie wavelength until the emission of gravitational waves takes over, i.e., $ 0.01 a_h\gg\lambdabar_\sigma$. The effect of dynamical friction on hardening the binary becomes less efficient when the individual encounters between the binary and the constituent particle of the surroundings (e.g., a field star) lasts longer than one orbital period~\cite{Begelman:1980vb}, 
which is given by $T(r)=2\pi r^{3/2}/\sqrt{GM}$. In the ULDM case, one can think of the coherence time, or equivalently, the lifetime of quasiparticles, as the encounter time between the binary and the individual quasiparticles. 
Hardening of SMBH binary, solely by dynamical friction, requires $T(0.01a_h)\gg \tau_c$, which implies that the binary encounters many quasiparticles during each revolution and therefore experiences dynamical friction. These two conditions, i.e., $ 0.01 a_h\gg\lambdabar_\sigma$ and $T(0.01a_h)\gg \tau_c$ are consistent with each other and point to a region of the parameter space within which the individual encounters between the quasiparticles and the SMBH binary may not be important. 

In general, close encounters between ULDM quasiparticles and a SMBH binary can play a role in hardening the binary. As we showed in subsection~\ref{subsec:ULDMSMBHencounter}, these encounters are able to extract energy from the binary in an efficient way which is comparable with field stars.
Regarding relaxation, according to Eq.~(\ref{eq:classrelax}), since the quasiparticles can be massive, we expect a reduction in the two-body relaxation time. The more accurate estimate of the two-body relaxation time in the halo of ULDM of mass $m_{\rm ULDM}$ is obtained as~\cite{2021ApJ...915...27B}:
\begin{equation}
    t_\text{relax}\simeq \frac{m_{\rm ULDM}^3\sigma^6}{G^2 \hbar^3\rho^2(r){\rm ln}(r/\lambdabar_\sigma)},
    \label{eq:ULDMrelax}    
\end{equation}
which is still too long to replenish the loss cone in a Hubble time. However, since the coherence time scale can be much shorter than the Hubble time, the relatively rapid appearance of quasiparticles can be understood as a way of replenishing the loss cone efficiently. 

Under reasonable assumptions, we can find the range of ULDM mass and SMBH mass corresponding to different regimes of interactions between a SMBH binary and ULDM halo. Without loss of generality and for simplicity, we focus on equal-mass SMBH binary, i.e., $M_1=M_2=M$.

To form a binary and start the hardening phase, SMBHs need to become close enough ($a\sim a_h$). As we mentioned in subsection~\ref{subsec:DFULDM}, when the de Broglie wavelength of ULDM is larger than the separation distance between the SMBHs, the dynamical friction is suppressed and even the formation of the binary may stall. To avoid this obstacle, we require that the binary forms outside of the solitonic core:  $a_h> \lambdabar_\sigma$. This leads to the following lower bound on the mass of the ULDM:
\begin{equation}
m_{\rm ULDM}>\frac{8\hbar\sigma}{GM}.
\label{eq:outsidecore}
\end{equation}
Below this limit, it is not clear, however, whether three-body encounters between individual quasiparticles and the almost formed SMBH binary can replace the weak dynamical friction and start the hardening phase. Even if they can, by lowering the mass of the ULDM, and increasing the size and consequently the mass of quasiparticles, we face another lower bound; large enough quasiparticles can inject energy into the binary and even break up the binary~\cite{Bar-Or:2018pxz}. To avoid heating up the binary with quasiparticles, the black holes must be heavier than the quasiparticles, i.e., $m_{\rm eff}(a_h)< M$, or in terms of the halo profile, $\rho(r)$, we demand
\begin{equation}
 \lambdabar_\sigma ^3 \rho(a_h)<M.
\end{equation}
The binary should encounter at least one quasiparticle during one complete revolution at the beginning of the hardening phase. To this end, the coherence time of the ULDM needs to exceed 
the orbital period, $T$. Therefore, we require  $T(a_h)< \tau_c$, which leads to the following upper bound on the mass of the ULDM:
\begin{equation}
   m_{\rm ULDM}<\frac{8\sqrt{2}}{\pi}\frac{\hbar \sigma}{GM},
\end{equation}
which is in strong tension with the bound in Eq.~(\ref{eq:outsidecore}).
After formation of a SMBH binary outside of the solitonic core ($a_h> \lambdabar_\sigma$), the quasiparticles may decrease the size of the binary by some factor, but a sizable reduction by almost two orders of magnitude brings the binary close to the solitonic core. The fate of the binary at this point depends on the interaction between the binary and the core. If the binary's tidal force pulls apart the core, black holes can keep losing energy to the quasiparticles and get closer. On the other hand, if the core persists, the binary may enter the core before getting small enough to start emitting GWs and therefore they may stall there.

From a theoretical point of view, the ability of a ULDM halo to extract energy from a secondary system can be related to the gravitational cooling process, which is basically emission of mass and energy to infinity~\cite{Seidel:1993zk,Guzman:2006yc,Schwabe:2016rze,Alvarez-Rios:2023cch}; a ULDM halo can absorb energy from the SMBH binary by adjusting its own profile (puffing up). In a recent paper, SMBH binaries inside ULDM halos have been studied numerically~\cite{2023arXiv231103412K}. As the initial condition of the simulation, the SMBH binary is assumed to be inside the solitonic core, with a separation of a few parsecs. 
This study does not explore the formation phase of the binary. For their chosen benchmark values of the black hole and ULDM masses, dynamical friction is already suppressed prior to formation of the binary, and the quasiparticles are heavier than the black holes, which can lead to injection of energy into the binary  black hole system. While the formation of the binary and its subsequent evolution to a separation of a few parsecs just by interacting with ULDM is not addressed and would likely be problematic, the study does demonstrate, for the parameters chosen, that a binary can lose energy to a halo efficiently via gravitational cooling. 
Some other possible complexities of the dynamics of a single SMBH inside a solitonic core, which may be generalized to two or more SMBHs, have been studied in Ref.~\cite{Wang:2021udl}.
The possible accretion of a solitonic core
by its resident black hole is also studied in Ref.~\cite{Cardoso:2022nzc} under the assumption of spherical symmetry. 

If we treat quasiparticles as stars, then, as  discussed in subsection~\ref{subsec:hardstar}, for a merger to occur on a reasonable time scale, an SMBH binary should interact with at least a collection of quasiparticles with overall mass of the order of the mass of the binary itself. This requires at least $M/m_{\rm eff}(a_h)$ encounters to happen in a Hubble time. In other words,
\begin{equation}
 \frac{\tau_{\rm Univ.}}{\tau_c}\gtrsim\frac{M}{m_{\rm eff}(a_h)},
\end{equation}
where $\tau_{\rm Univ.}$ denotes the age of the Universe.

By increasing the mass of ULDM, at some point the de Broglie wavelength becomes comparable with the Schwarzschild radius of the SMBHs, $R_S=2GM/c^2$. In the limit $\lambdabar_\sigma \lesssim R_S$, the minimum impact parameter, or equivalently, the softening scale, coincides with $R_S$ and therefore ULDM can be treated as cold DM. 
In this case, the final parsec problem persists, as long as the system is spherically symmetric. This condition provides an upper bound on the mass of ULDM, given by:
\begin{equation}
 m_{\rm ULDM}\lesssim \frac{\hbar c^2}{2GM\sigma}.  
\end{equation}
Deviations from spherical symmetry may lead to efficient refilling of the cold DM loss cone. However, it remains unclear whether constant interactions between a binary and cold DM particles can give rise to a merger in a Hubble time. We leave an accurate analysis of this case for future work.

\section{Results}\label{sec:results}

Here we present our results.  First, we identify regions of the $(M_{\rm BH},m_{\rm ULDM})$ parameter space within which the SMBH binary can merge in less than a Hubble time, then we consider the interplay between ULDM and stars in solving the final parsec problem.

\subsection{SMBH Binary in ULDM}

\begin{figure}[t]
  \centering
    \includegraphics[width=0.45\textwidth]{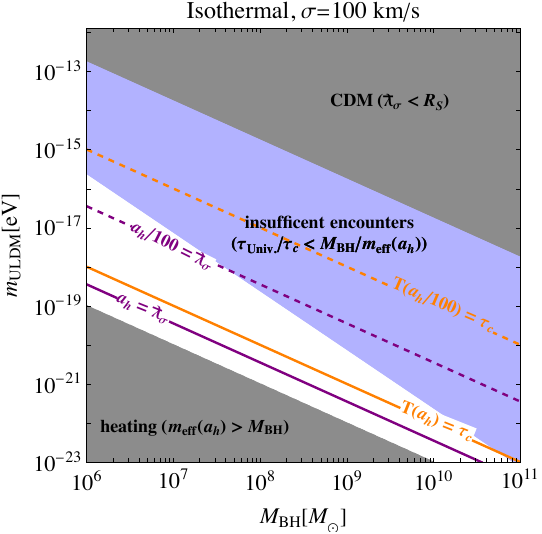}
    \includegraphics[width=0.45\textwidth ]{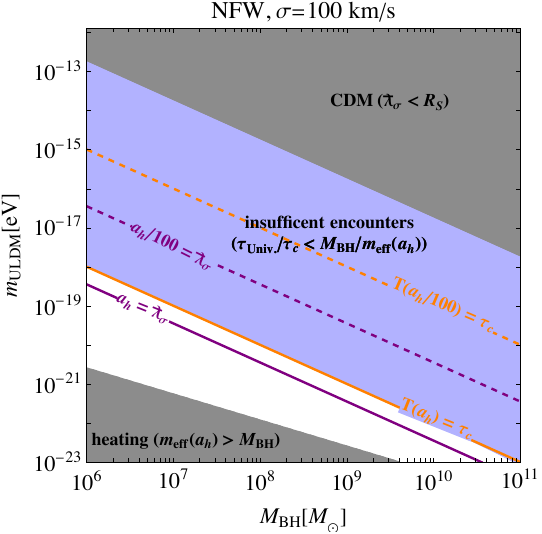}
  \caption{Different regions of the $(M_{\rm BH},m_{\rm ULDM})$ parameter space correspond to different regimes
of interactions between a SMBH binary of equal mass and ULDM halo. Left (Right) panel corresponds to the isothermal (NFW) profile with $\sigma=100\,{\rm km/s}$.
In each panel of, the gray shaded region at the bottom shows where the effective mass of quasiparticles is larger than the mass of SMBHs and binary formation and merger can stall due to absorption of energy from halo. Within the gray shaded region at the top, the de Broglie wavelength of ULDM is smaller than  the Schwarzschild radius of the SMBHs and ULDM can be treated as cold particle-like DM. Between these two gray regions, ULDM wave nature can play an important role in merger of SMBH binary. 
The purple solid (dashed) line corresponds to contour of $a_h=\lambdabar_\sigma$ ( $a_h/100=\lambdabar_\sigma$).
The orange solid (dashed) line marks $T(a_h)=\tau_c$ ($T(a_h/100)=\tau_c$). Below the orange solid line, which includes $a_h=\lambdabar_\sigma$, we have $T(a_h)<\tau_c$ and therefore individual encounters are important. Similarly, below the orange dashed line, which contains $a_h/100=\lambdabar_\sigma$, we have $T(a_h/100)<\tau_c$ and therefore dynamical friction cannot describe the interactions fully. 
Within the blue shaded region in Fig.~\ref{fig:parameterspace1}, SMBH binary does not encounter a sufficient number of quasiparticles.}
  \label{fig:parameterspace1}
\end{figure}

In Fig.~\ref{fig:parameterspace1}, we present the regions of the $(M_{\rm BH},m_{\rm ULDM})$ parameter space that correspond to different regimes of interactions between a SMBH binary of equal mass and ULDM halo, as discussed in detail 
above. In the left panel of Fig.~\ref{fig:parameterspace1} we assume an isothermal profile for the halo outside of the solitonic core, given by
\begin{equation}
    \rho_{\rm Iso}(r)=\frac{\sigma^2}{2\pi G r^2},
\end{equation}
and in the right panel we assume an NFW profile, given by~\cite{Navarro:1996gj}
\begin{equation}
    \rho_{\rm NFW}(r)=\rho_s\frac{r_s}{r}\left(1+\frac{r}{r_s}\right)^{-2}.
\end{equation}
Here, $\sigma$ is the dispersion velocity and is assumed to be equal to $100\,{\rm km}/{\rm s}$ for both profiles\footnote{We also investigated $\sigma=200\,{\rm km}/{\rm s}$, and the results are not qualitatively different. For brevity, we focus here on $\sigma=100\,{\rm km}/{\rm s}$.}, $\rho_s=0.184\,{\rm GeV}/{\rm cm}^{3}$, and $r_s=24.42\,{\rm kpc}$~\cite{Cirelli:2010xx}.
In each panel of Fig.~\ref{fig:parameterspace1}, the gray shaded region at the bottom is where the effective mass of quasiparticles is larger than the mass of SMBHs; in this region, as we discussed in subsection~\ref{subsection:obpo}, quasiparticles can heat up the binary and prevent the merger. Within the gray shaded region at the top, the de Broglie wavelength of ULDM becomes smaller than  the Schwarzschild radius of the SMBHs so that ULDM can be treated as cold particle-like DM, and therefore the wave nature of ULDM cannot alleviate the final parsec problem.
Between these two gray regions, the wave nature of ULDM can play an important role in the mergers of SMBH binaries. A combination of dynamical friction and individual encounters between quasiparticles and the binary can harden the binary, allowing it to reach the stage where emission of gravitational waves takes over and the two black holes merge. 
Since transitioning from efficient dynamical friction to individual encounters does not occur sharply, we compare the coherence time (or equivalently the lifetime of quasiparticles) to the orbital period to to distinguish the two regimes: roughly speaking, if the orbital period of the binary is longer than the coherence time, the binary interacts with several quasiparticles and therefore experiences dynamical friction, otherwise individual encounters play the dominant role.  

To demonstrate the importance of individual encounters with quasiparticles, in Fig.~\ref{fig:parameterspace1}, we display contours of $a_h=\lambdabar_\sigma$ (purple solid) and $a_h/100=\lambdabar_\sigma$(purple dashed). Above the purple solid contour, the size of the binary at formation is larger than $\lambdabar_\sigma$, and above the purple dashed contour, the binary is larger than $100\lambdabar_\sigma$ when it forms.  
We also compare the orbital period with the coherence time at binary formation: along the orange solid and dashed lines, $T(a_h)=\tau_c$ and $T(a_h/100)=\tau_c$, respectively.
Below the orange solid line, which includes $a_h=\lambdabar_\sigma$, we have $T(a_h)<\tau_c$ and therefore individual encounters are important. Similarly, below the orange dashed line, which contains $a_h/100=\lambdabar_\sigma$, we have $T(a_h/100)<\tau_c$ and therefore dynamical friction cannot describe the interactions fully.
If individual encounters between quasiparticles and the SMBH binary can extract energy efficiently from it, a sufficient number of encounters between the binary and quasiparticles can lead to a merger in under a Hubble time. Within the blue shaded region in Fig.~\ref{fig:parameterspace1}, the SMBH binary does not encounter a sufficient number of quasiparticles.

\subsection{SMBH Binary in ULDM, Inclusion of Stars}
So far, we reviewed the evolution of SMBH binaries due to solely stellar dynamics, and we also discussed the possible effects of ULDM on SMBH binaries in the absence of stars. 
A more realistic and interesting case includes both stars and ULDM, and the role that the dynamics between them plays in the merger of SMBH binaries. 
One of the proposed solutions to the final parsec problem is refilling the loss cone by scattering stars off the massive perturbers in galaxies~\cite{Perets:2006bz,Perets:2007nc}. A steady state population of massive perturbers such as molecular clouds, or globular clusters can capture scattered stars and return them back to the central region of galaxies.
In this regard, ULDM quasiparticles are unique astrophysical objects in that they naturally appear throughout the halo and are particularly effective as massive perturbers near the center of a galaxy, where the local density is high.

Massive perturbers can accelerate the stellar relaxation locally by several orders of magnitude~\cite{Perets:2006bz,Perets:2007nc}. Basically a population of massive perturbers that are much more massive than individual stars can easily dominate the relaxation process by gravitational scattering within the region that contains them. 
Consider a test star of mass $M_\star$ moving with relative velocity $v$ at a distance of the order of the capture radius, $r_c\sim GM_p/v^2$ of a population of perturbers with mass $M_p$ and number density $n_p$.  The rate of encounters between the test star and massive perturbers is $n_p v \sigma_{p\star}$ where $\sigma_{p\star}\sim r_c^2$ is the capture cross-section. Therefore the relaxation time due to scattering off the massive perturbers is given by $t_r\sim (G^2/v^3) n_p M_p^2$. 
Including all the encounter distances by integrating over them decreases the relaxation time by a Coulomb logarithm factor which in general depends on the size of the system and the size of the perturbers. It has been shown that the ratio of the second moments of the mass distributions of massive perturbers and stars roughly determines the amount of relaxation acceleration due to perturbers~\cite{Perets:2006bz,Perets:2007nc}. In other words, we have
\begin{equation}
    \frac{t_{r,\star}}{t_{r,p}}\sim \mu_2\equiv \frac{n_p M_p^2}{n_\star M_\star^2},
\end{equation}
where $t_{r,\star}$ is the relaxation time scale due to stellar two-body scattering, $t_{r,p}$ is the relaxation time scale due to scattering of stars off the massive perturbers, and $n_{(p,\star)}$ and $M_{(p,\star)}$ are the number density and mass of the perturbers and stars, respectively.

In astrophysical contexts, massive perturbers such as giant molecular clouds or globular clusters,
located within a few hundred parsecs of a galactic center, 
can replenish the loss cone efficiently \cite{Perets:2006bz}. Massive perturbers in the mass range $\sim 10^4-10^7\,M_\odot$, with $\mu_2\sim 10^5-10^7$, can lead to a SMBH binary merger within a Hubble time over a wide range of black hole masses~\cite{Perets:2007nc}. 

By the same reasoning,  quasiparticles in a ULDM halo, within the right range of mass and number density, can scatter stars back to the loss cone. The big difference between quasiparticles in a ULDM halo and traditional massive perturbers is their ubiquity; provided that ULDM constitutes the dominant component of DM in the Universe, then the quasiparticles appear in every halo and consequently in every galaxy, while traditional perturbers lack this universality.  For instance, while giant molecular clouds are common in the disks of spiral galaxies, they do not survive in elliptical galaxies as a result of a history of major mergers.

To find out in which part of the parameter space quasiparticles can enhance the relaxation rate noticeably, we evaluate the parameter $\mu_2$ in a region of the size of $a_h$ enclosed by the SMBH binary. Hence, we obtain
\begin{equation}
    \mu_2= \frac{n_{\rm dB} m_{\rm eff}^2}{n_\star M_\star^2}\sim \frac{ m_{\rm eff}}{M_\star}\frac{\int_0^{a_h} dr\, r^2 \rho_{\rm ULDM}(r)}{\int_0^{a_h} dr\, r^2\rho_\star(r)},
\end{equation}
where $n_{\rm dB}$ is the number density of quasiparticles. For the stellar density we assume an isothermal distribution,
\begin{equation}
    \rho_\star(r)=\frac{\sigma^2}{2\pi G r^2},
\end{equation}
and for the mass of quasiparticles we use the local density profile of ULDM to evaluate the mass enclosed in a de Broglie volume, following Eq.~(\ref{eq:effmass}).
To be more conservative, we evaluate the mass of quasiparticles at $r=a_h$.
While heavy enough quasiparticles can act like massive perturbers, they can also heat up the SMBH binary. So the accelerated relaxation rate due to quasiparticles is efficient as long as the quasiparticles are not heavier than the black holes, i.e.~$m_{\rm eff}<M_{\rm BH}$. When the mass of the quasiparticles become comparable to the black hole mass or larger, one needs to analyse the system more carefully to find out the overall effect of the quasiparticles on the binary, which will be a combination of heating it up and refilling the loss cone with stars. 
The quasiparticles are most effective when their lifetime becomes comparable to the orbital period of the SMBH binary, i.e.~$T(a_h)\sim \tau_c$.

\begin{figure}[htbp]
  \centering
    \includegraphics[width=0.45\textwidth]{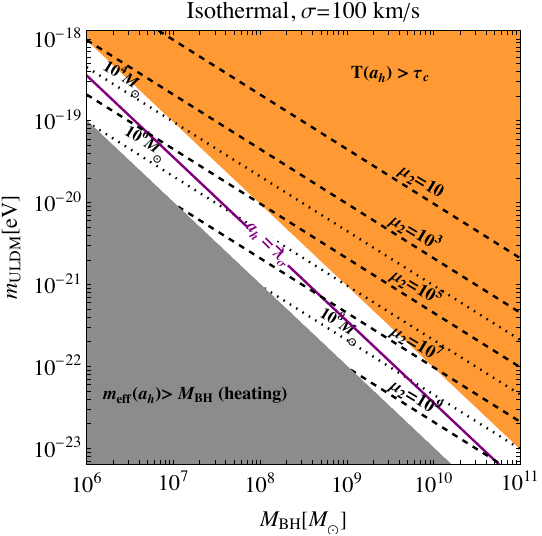}
    \includegraphics[width=0.45\textwidth]{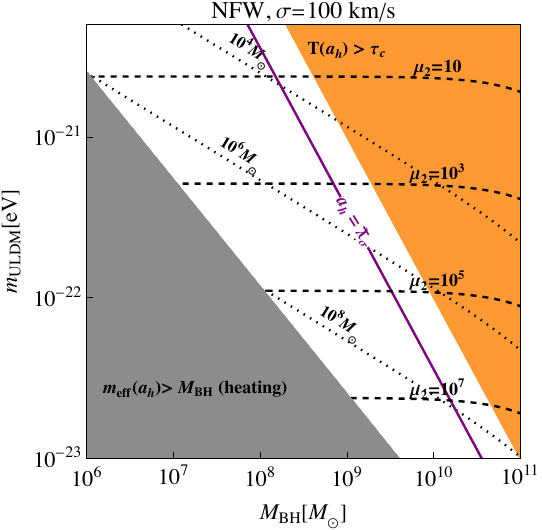}
  \caption{The estimated local enhancement in stellar relaxation rate in galaxy due to quasiparticles of ULDM. Left (Right) panel corresponds to the isothermal (NFW) profile with $\sigma=100\,{\rm km/s}$.
In each panel, the gray shaded region shows where the effective mass of quasiparticles is larger than the mass of SMBH.
The black dashed contours display $\mu_2$ parameter which represents the local enhancement in stellar relaxation rate. Within the orange shaded region, the lifetime of quasiparticles is less than the orbital period and the impact of them is not maximal. The black dotted contours show the mass of a quasiparticle at $a_h$.}
\label{fig:massiveperturber}
\end{figure}

The black dashed contours in Fig.~\ref{fig:massiveperturber} depict the $\mu_2$ parameter which represents the local enhancement in the stellar relaxation rate in the galaxy. The left (right) panel corresponds to an isothermal (NFW) profile with $\sigma=100\,{\rm km/s}$. Within the orange shaded region, the lifetime of quasiparticles is less than the orbital period of the binary, and their impact in scattering stars back to the loss cone is not maximal. The black dotted contours show the mass of a quasiparticle at $a_h$.

Here, we comment briefly on the effect of traditional cold DM subhalos on the final parsec problem:  One might imagine that a massive cold DM subhalo could serve as the massive perturber, effecting the phase space adequately that the loss cone can be replenished. As of the writing of this paper, we find no evidence in the literature that cold DM subhalos would be effective as massive perturbers.  For a virialized halo, the virial radius for a subhalo of the size required to serve as a massive perturber would exceed the size of the binary black hole system by several orders of magnitude (though cold DM subhalos may have a central density cusp).  Furthermore, the number density of large enough subhalos in the inner $\sim 100$ pc is expected to be so low~\cite{Stref:2019wjv} that that it seems extremely unlikely that cold DM subhalos could serve as effective massive perturbers.  Nonetheless, detailed simulations could shed further light on this possibility.

\section{Summary}\label{sec:discussandconclude}
In this paper we have explored the evolution of SMBH binaries in galaxies with ULDM halos. As a result of the wave nature of ULDM, and the consequent interference of its excited modes, density fluctuations in the DM halo appear as massive quasiparticles. The lifetime of these quasiparticles is defined by their coherence time, which is much shorter than the Hubble time.  To exploit the wave nature of ULDM in resolving the final parsec problem, the de Broglie wavelength of ULDM must be longer than the Schwarzschild radius of the SMBHs. To avoid suppression of dynamical friction from ULDM and achieve a successful formation of the binary, SMBH should start the hardening phase outside of the solitonic core. When the mass of the quasiparticles are comparable to the mass of SMBHs or larger, they heat up the binary instead of extracting energy from it. By applying these considerations, we have mapped out the mass ranges of ULDM and SMBHs within which dynamical friction from ULDM together with individual encounters between quasiparticles and the SMBH binary may lead to a merger of the binary in under a Hubble time. Because the lifetime of the quasiparticles is typically longer than the orbital period of the binary at the beginning of the hardening phase, the role of individual encounters between quasiparticles and the binary in shrinking the size of the binary becomes important.  On the basis of preliminary simulations, we predict that quasiparticles can indeed be as efficient as field stars in extracting energy from the binary during individual encounters; ULDM quasiparticles have a key advantage over field stars in that their appearance is inevitable and there is therefore no need to replenish the loss cone.

By adding stars to this scenario, ULDM quasiparticles can resolve the final parsec problem in a completely novel way; they can act as universal massive perturbers to enhance the local stellar relaxation rate by scattering stars back into the loss cone. The importance of this new class of massive perturbers can be understood better in connection with elliptical galaxies, which lack regular massive perturbers such as molecular clouds as a result of a history of major mergers. 

The main goal of this study is to explore the role of ULDM in the merger of SMBH binaries. We considered a parameter space of DM properties, halo structure, and black hole masses to identify conditions where ULDM can facilitate the formation of binaries as well as drive their subsequent evolution. Scenarios for ULDM as a catalyst for mergers --- either through direct interactions or by stirring up the phase space of stars --- are promising avenues for future work on SMBH binaries as the gravitational universe opens to us.

\acknowledgments
We are grateful to Lam Hui for 
encouragement on this project and Mustafa A. Amin, Mike Boylan-Kolchin, Yao-Yuan Mao, and Paul R. Shapiro for extremely helpful conversations, and to the Department of Physics and Astronomy, University of Utah, where part of the work was completed. The work of B.S.E.~is supported in part by DOE grant DE-SC0022021.  The work of P.S.~is supported in part by NSF grant PHY-2014075.
Aspects of this research were made possible thanks to a supercomputing allocation on NASA's Discover cluster under NASA grant 80NSSC23K0252.

\appendix*
\section{Numerical approach}\label{appx:num}

To assess the outcome of an ULDM quasiparticle interacting with a SMBH binary, we devoloped a simple code to solve the Schr\"{o}dinger equation with a binary black hole potential, along with a solution to the binary orbit in the presence of ULDM mass distribution. The computational domain is triply-periodic, spanning a cube with length $L = 8$~pc on each side, and centered at the origin of a rectilinear coordinate system with coordinates ($x,y,z$). The binary, represented as classical, Newtonian $n$-body particles, has a total mass $M_1+M_2 = 10^7~M_\odot$, a center of mass at the origin, and an initial orbital separation of $a_0 = 1$~pc. The binary's phase and orientation (direction of its angular momentum vector) are chosen at random.  The wavefunction $\psi$ is defined with real and imaginary parts on a regular grid with $N_L^3 = 1000^3$ points. Its mass, $M_\psi$ is 1\%\ of the total binary mass, while the mass of the ULDM particles that compose it is $m \approx 3\times 10^{-19}$~eV. We choose the quasiparticle's packet size and de Brogie wavelength so that initially $R = \lambda = 0.2$~pc, and its initial central location is near the domain edge at $(-L/2,0,0)$ approximately 4~pc from the binary. Its propagation speed is $v_0 \approx 170$~km/s in the $+x$ direction toward the origin, consistent with a quasiparticle that had a speed of 100~/km/s, representative of the ULDM bulk, but that has fallen into the gravitational well of the black hole binary.

We step the quasiparticle-binary system forward in time using a finite-difference PDE solver for $\psi$ and an ODE solver for the binary partners. The grid-based finite differences approximate space and time derivatives in the Schr\"{o}dinger equation; the spatial differences depend on the grid spacing, $\Delta L = 0.008$~pc, while the time step, $\Delta t$, is bounded by $\Delta L^2/4\hbar$, per the Courant condition for numerical stability. A leap-frog scheme advances the real and imaginary parts of $\psi$ in time, with $\text{Re}\psi$ and $\text{Im} \psi$ staggered by half a time step. The particle trajectories are calculated with a $6^\text{th}$-order symplectic integrator \cite{yoshida1990}, with time steps that are smaller than the finite-difference steps so the interaction between the field and the black holes is updated with every change to $\psi$. The forces between the black holes and the field are softened on a scale of 0.016~pc, twice the grid spacing. 

Our code, written in C++ and using OpenMP for parallelization, is available upon request to the authors.

We apply this evolution algorithm to a suite of initial conditions. Each simulation is evolved for a time $t = 1.5 L/v_0$, approximately twice the binary orbital period. After the start of a simulation, and after the quasiparticle had traveled approximately a quarter of the way to the black hole pair, an algorithm is turned on that strongly attenuates the wavefunction in a boundary layer of about $\ell = 0.02$~pc width from each face of the cubic computational domain. There, $\psi$ is artificially suppressed each time step, reduced by a factor of $\exp(-\Delta r/\ell)$ where $\Delta r$ is the distance between a grid location and the nearest edge of the domain. Otherwise, wavefunction components that are scattered outward would wrap around in the triply periodic computational domain to repeatedly scatter off the binary. There are elegant ways of imposing radiative boundary conditions, but this method is simple, numerically stable, and does what we want: it allows outgoing waves to effectively leave the computational domain. 

Our final suite of simulations consists of 200 runs. Each run, with 1000$^3$ grid points, requires approximately 10 hours on a 40-core computational node. Fig.~\ref{fig:fuzzysim} shows example snapshots of the quasiparticle-binary interaction in one simulation. In the Figure, the wavefunction scatters off of one binary partner only to diffract off of the other black hole as it swings around in its orbit. Much of the wavefunction is lost to regions outside the computational domain after this first encounter; the rest lingers near the binary and will eventually be ejected as well. Presumably, some small fraction of the quasiparticle is bound by the binary and will eventually accrete onto the black holes. 

\bibliography{dynfric}{}

\begin{thebibliography}{64}%
\makeatletter
\providecommand \@ifxundefined [1]{%
 \@ifx{#1\undefined}
}%
\providecommand \@ifnum [1]{%
 \ifnum #1\expandafter \@firstoftwo
 \else \expandafter \@secondoftwo
 \fi
}%
\providecommand \@ifx [1]{%
 \ifx #1\expandafter \@firstoftwo
 \else \expandafter \@secondoftwo
 \fi
}%
\providecommand \natexlab [1]{#1}%
\providecommand \enquote  [1]{``#1''}%
\providecommand \bibnamefont  [1]{#1}%
\providecommand \bibfnamefont [1]{#1}%
\providecommand \citenamefont [1]{#1}%
\providecommand \href@noop [0]{\@secondoftwo}%
\providecommand \href [0]{\begingroup \@sanitize@url \@href}%
\providecommand \@href[1]{\@@startlink{#1}\@@href}%
\providecommand \@@href[1]{\endgroup#1\@@endlink}%
\providecommand \@sanitize@url [0]{\catcode `\\12\catcode `\$12\catcode
  `\&12\catcode `\#12\catcode `\^12\catcode `\_12\catcode `\%12\relax}%
\providecommand \@@startlink[1]{}%
\providecommand \@@endlink[0]{}%
\providecommand \url  [0]{\begingroup\@sanitize@url \@url }%
\providecommand \@url [1]{\endgroup\@href {#1}{\urlprefix }}%
\providecommand \urlprefix  [0]{URL }%
\providecommand \Eprint [0]{\href }%
\providecommand \doibase [0]{http://dx.doi.org/}%
\providecommand \selectlanguage [0]{\@gobble}%
\providecommand \bibinfo  [0]{\@secondoftwo}%
\providecommand \bibfield  [0]{\@secondoftwo}%
\providecommand \translation [1]{[#1]}%
\providecommand \BibitemOpen [0]{}%
\providecommand \bibitemStop [0]{}%
\providecommand \bibitemNoStop [0]{.\EOS\space}%
\providecommand \EOS [0]{\spacefactor3000\relax}%
\providecommand \BibitemShut  [1]{\csname bibitem#1\endcsname}%
\let\auto@bib@innerbib\@empty
\bibitem [{\citenamefont {Richstone}\ \emph {et~al.}(1998)\citenamefont
  {Richstone} \emph {et~al.}}]{Richstone:1998ky}%
  \BibitemOpen
  \bibfield  {author} {\bibinfo {author} {\bibfnamefont {D.}~\bibnamefont
  {Richstone}} \emph {et~al.},\ }\href@noop {} {\bibfield  {journal} {\bibinfo
  {journal} {Nature}\ }\textbf {\bibinfo {volume} {395}},\ \bibinfo {pages}
  {A14} (\bibinfo {year} {1998})},\ \Eprint
  {http://arxiv.org/abs/astro-ph/9810378} {arXiv:astro-ph/9810378} \BibitemShut
  {NoStop}%
\bibitem [{\citenamefont {Kormendy}\ and\ \citenamefont
  {Ho}(2013)}]{Kormendy:2013dxa}%
  \BibitemOpen
  \bibfield  {author} {\bibinfo {author} {\bibfnamefont {J.}~\bibnamefont
  {Kormendy}}\ and\ \bibinfo {author} {\bibfnamefont {L.~C.}\ \bibnamefont
  {Ho}},\ }\href {\doibase 10.1146/annurev-astro-082708-101811} {\bibfield
  {journal} {\bibinfo  {journal} {Ann. Rev. Astron. Astrophys.}\ }\textbf
  {\bibinfo {volume} {51}},\ \bibinfo {pages} {511} (\bibinfo {year} {2013})},\
  \Eprint {http://arxiv.org/abs/1304.7762} {arXiv:1304.7762 [astro-ph.CO]}
  \BibitemShut {NoStop}%
\bibitem [{\citenamefont {Milosavljevic}\ and\ \citenamefont
  {Merritt}(2003{\natexlab{a}})}]{Milosavljevic:2002bn}%
  \BibitemOpen
  \bibfield  {author} {\bibinfo {author} {\bibfnamefont {M.}~\bibnamefont
  {Milosavljevic}}\ and\ \bibinfo {author} {\bibfnamefont {D.}~\bibnamefont
  {Merritt}},\ }\href {\doibase 10.1086/378086} {\bibfield  {journal} {\bibinfo
   {journal} {Astrophys. J.}\ }\textbf {\bibinfo {volume} {596}},\ \bibinfo
  {pages} {860} (\bibinfo {year} {2003}{\natexlab{a}})},\ \Eprint
  {http://arxiv.org/abs/astro-ph/0212459} {arXiv:astro-ph/0212459} \BibitemShut
  {NoStop}%
\bibitem [{\citenamefont {Yu}(2002)}]{Yu:2001xp}%
  \BibitemOpen
  \bibfield  {author} {\bibinfo {author} {\bibfnamefont {Q.}~\bibnamefont
  {Yu}},\ }\href {\doibase 10.1046/j.1365-8711.2002.05242.x} {\bibfield
  {journal} {\bibinfo  {journal} {Mon. Not. Roy. Astron. Soc.}\ }\textbf
  {\bibinfo {volume} {331}},\ \bibinfo {pages} {935} (\bibinfo {year}
  {2002})},\ \Eprint {http://arxiv.org/abs/astro-ph/0109530}
  {arXiv:astro-ph/0109530} \BibitemShut {NoStop}%
\bibitem [{\citenamefont {Agazie}\ \emph
  {et~al.}(2023{\natexlab{a}})\citenamefont {Agazie} \emph
  {et~al.}}]{NANOGrav:2023gor}%
  \BibitemOpen
  \bibfield  {author} {\bibinfo {author} {\bibfnamefont {G.}~\bibnamefont
  {Agazie}} \emph {et~al.} (\bibinfo {collaboration} {NANOGrav}),\ }\href
  {\doibase 10.3847/2041-8213/acdac6} {\bibfield  {journal} {\bibinfo
  {journal} {Astrophys. J. Lett.}\ }\textbf {\bibinfo {volume} {951}},\
  \bibinfo {pages} {L8} (\bibinfo {year} {2023}{\natexlab{a}})},\ \Eprint
  {http://arxiv.org/abs/2306.16213} {arXiv:2306.16213 [astro-ph.HE]}
  \BibitemShut {NoStop}%
\bibitem [{\citenamefont {{Antoniadis}}\ \emph {et~al.}(2023)\citenamefont
  {{Antoniadis}} \emph {et~al.}}]{EPTA:2023fyk}%
  \BibitemOpen
  \bibfield  {author} {\bibinfo {author} {\bibfnamefont {J.}~\bibnamefont
  {{Antoniadis}}} \emph {et~al.} (\bibinfo {collaboration} {EPTA}),\ }\href
  {\doibase 10.1051/0004-6361/202346844} {\bibfield  {journal} {\bibinfo
  {journal} {Astron. Astrophys.}\ }\textbf {\bibinfo {volume} {678}},\ \bibinfo
  {eid} {A50} (\bibinfo {year} {2023})},\ \Eprint
  {http://arxiv.org/abs/2306.16214} {arXiv:2306.16214 [astro-ph.HE]}
  \BibitemShut {NoStop}%
\bibitem [{\citenamefont {Reardon}\ \emph {et~al.}(2023)\citenamefont {Reardon}
  \emph {et~al.}}]{Reardon:2023gzh}%
  \BibitemOpen
  \bibfield  {author} {\bibinfo {author} {\bibfnamefont {D.~J.}\ \bibnamefont
  {Reardon}} \emph {et~al.},\ }\href {\doibase 10.3847/2041-8213/acdd02}
  {\bibfield  {journal} {\bibinfo  {journal} {Astrophys. J. Lett.}\ }\textbf
  {\bibinfo {volume} {951}},\ \bibinfo {pages} {L6} (\bibinfo {year} {2023})},\
  \Eprint {http://arxiv.org/abs/2306.16215} {arXiv:2306.16215 [astro-ph.HE]}
  \BibitemShut {NoStop}%
\bibitem [{\citenamefont {Xu}\ \emph {et~al.}(2023)\citenamefont {Xu} \emph
  {et~al.}}]{Xu:2023wog}%
  \BibitemOpen
  \bibfield  {author} {\bibinfo {author} {\bibfnamefont {H.}~\bibnamefont {Xu}}
  \emph {et~al.},\ }\href {\doibase 10.1088/1674-4527/acdfa5} {\bibfield
  {journal} {\bibinfo  {journal} {Res. Astron. Astrophys.}\ }\textbf {\bibinfo
  {volume} {23}},\ \bibinfo {pages} {075024} (\bibinfo {year} {2023})},\
  \Eprint {http://arxiv.org/abs/2306.16216} {arXiv:2306.16216 [astro-ph.HE]}
  \BibitemShut {NoStop}%
\bibitem [{\citenamefont {Afzal}\ \emph {et~al.}(2023)\citenamefont {Afzal}
  \emph {et~al.}}]{NANOGrav:2023hvm}%
  \BibitemOpen
  \bibfield  {author} {\bibinfo {author} {\bibfnamefont {A.}~\bibnamefont
  {Afzal}} \emph {et~al.} (\bibinfo {collaboration} {NANOGrav}),\ }\href
  {\doibase 10.3847/2041-8213/acdc91} {\bibfield  {journal} {\bibinfo
  {journal} {Astrophys. J. Lett.}\ }\textbf {\bibinfo {volume} {951}},\
  \bibinfo {pages} {L11} (\bibinfo {year} {2023})},\ \Eprint
  {http://arxiv.org/abs/2306.16219} {arXiv:2306.16219 [astro-ph.HE]}
  \BibitemShut {NoStop}%
\bibitem [{\citenamefont {Agazie}\ \emph
  {et~al.}(2023{\natexlab{b}})\citenamefont {Agazie} \emph
  {et~al.}}]{NANOGrav:2023hfp}%
  \BibitemOpen
  \bibfield  {author} {\bibinfo {author} {\bibfnamefont {G.}~\bibnamefont
  {Agazie}} \emph {et~al.} (\bibinfo {collaboration} {NANOGrav}),\ }\href
  {\doibase 10.3847/2041-8213/ace18b} {\bibfield  {journal} {\bibinfo
  {journal} {Astrophys. J. Lett.}\ }\textbf {\bibinfo {volume} {952}},\
  \bibinfo {pages} {L37} (\bibinfo {year} {2023}{\natexlab{b}})},\ \Eprint
  {http://arxiv.org/abs/2306.16220} {arXiv:2306.16220 [astro-ph.HE]}
  \BibitemShut {NoStop}%
\bibitem [{\citenamefont {Milosavljevic}\ and\ \citenamefont
  {Merritt}(2003{\natexlab{b}})}]{Milosavljevic:2002ht}%
  \BibitemOpen
  \bibfield  {author} {\bibinfo {author} {\bibfnamefont {M.}~\bibnamefont
  {Milosavljevic}}\ and\ \bibinfo {author} {\bibfnamefont {D.}~\bibnamefont
  {Merritt}},\ }\href {\doibase 10.1063/1.1629432} {\bibfield  {journal}
  {\bibinfo  {journal} {AIP Conf. Proc.}\ }\textbf {\bibinfo {volume} {686}},\
  \bibinfo {pages} {201} (\bibinfo {year} {2003}{\natexlab{b}})},\ \Eprint
  {http://arxiv.org/abs/astro-ph/0212270} {arXiv:astro-ph/0212270} \BibitemShut
  {NoStop}%
\bibitem [{\citenamefont {{Barausse}}\ and\ \citenamefont
  {{Lapi}}(2021)}]{Barausse:2020kjy}%
  \BibitemOpen
  \bibfield  {author} {\bibinfo {author} {\bibfnamefont {E.}~\bibnamefont
  {{Barausse}}}\ and\ \bibinfo {author} {\bibfnamefont {A.}~\bibnamefont
  {{Lapi}}},\ }in\ \href {\doibase 10.1007/978-981-15-4702-7_18-1} {\emph
  {\bibinfo {booktitle} {Handbook of Gravitational Wave Astronomy}}}\ (\bibinfo
   {publisher} {Springer Nature},\ \bibinfo {year} {2021})\ p.~\bibinfo {pages}
  {18}\BibitemShut {NoStop}%
\bibitem [{\citenamefont {Quinlan}\ and\ \citenamefont
  {Hernquist}(1997)}]{Quinlan:1997qe}%
  \BibitemOpen
  \bibfield  {author} {\bibinfo {author} {\bibfnamefont {G.~D.}\ \bibnamefont
  {Quinlan}}\ and\ \bibinfo {author} {\bibfnamefont {L.}~\bibnamefont
  {Hernquist}},\ }\href {\doibase 10.1016/S1384-1076(97)00039-0} {\bibfield
  {journal} {\bibinfo  {journal} {New Astron.}\ }\textbf {\bibinfo {volume}
  {2}},\ \bibinfo {pages} {533} (\bibinfo {year} {1997})},\ \Eprint
  {http://arxiv.org/abs/astro-ph/9706298} {arXiv:astro-ph/9706298} \BibitemShut
  {NoStop}%
\bibitem [{\citenamefont
  {{Holley-Bockelmann}}(2006)}]{Holley-Bockelmann:2006gbs}%
  \BibitemOpen
  \bibfield  {author} {\bibinfo {author} {\bibfnamefont {J.}~\bibnamefont
  {{Holley-Bockelmann}}},\ }\href@noop {} {\enquote {\bibinfo {title}
  {{Implications of Stellar Interactions with Supermassive Black Holes}},}\
  }\bibinfo {howpublished} {Presented at the KITP Program: Physics of Galactic
  Nuclei, Jul 20, 2006, Kavli Institute for Theoretical Physics, University of
  California, Santa Barbara, id.4} (\bibinfo {year} {2006})\BibitemShut
  {NoStop}%
\bibitem [{\citenamefont {Berczik}\ \emph {et~al.}(2006)\citenamefont
  {Berczik}, \citenamefont {Merritt}, \citenamefont {Spurzem},\ and\
  \citenamefont {Bischof}}]{Berczik:2006tz}%
  \BibitemOpen
  \bibfield  {author} {\bibinfo {author} {\bibfnamefont {P.}~\bibnamefont
  {Berczik}}, \bibinfo {author} {\bibfnamefont {D.}~\bibnamefont {Merritt}},
  \bibinfo {author} {\bibfnamefont {R.}~\bibnamefont {Spurzem}}, \ and\
  \bibinfo {author} {\bibfnamefont {H.-P.}\ \bibnamefont {Bischof}},\ }\href
  {\doibase 10.1086/504426} {\bibfield  {journal} {\bibinfo  {journal}
  {Astrophys. J. Lett.}\ }\textbf {\bibinfo {volume} {642}},\ \bibinfo {pages}
  {L21} (\bibinfo {year} {2006})},\ \Eprint
  {http://arxiv.org/abs/astro-ph/0601698} {arXiv:astro-ph/0601698} \BibitemShut
  {NoStop}%
\bibitem [{\citenamefont {{Khan}}\ \emph {et~al.}(2011)\citenamefont {{Khan}},
  \citenamefont {{Just}},\ and\ \citenamefont
  {{Merritt}}}]{2011ApJ...732...89K}%
  \BibitemOpen
  \bibfield  {author} {\bibinfo {author} {\bibfnamefont {F.~M.}\ \bibnamefont
  {{Khan}}}, \bibinfo {author} {\bibfnamefont {A.}~\bibnamefont {{Just}}}, \
  and\ \bibinfo {author} {\bibfnamefont {D.}~\bibnamefont {{Merritt}}},\ }\href
  {\doibase 10.1088/0004-637X/732/2/89} {\bibfield  {journal} {\bibinfo
  {journal} {\apj}\ }\textbf {\bibinfo {volume} {732}},\ \bibinfo {eid} {89}
  (\bibinfo {year} {2011})},\ \Eprint {http://arxiv.org/abs/1103.0272}
  {arXiv:1103.0272 [astro-ph.CO]} \BibitemShut {NoStop}%
\bibitem [{\citenamefont {{Holley-Bockelmann}}\ and\ \citenamefont
  {{Khan}}(2015)}]{2015ApJ...810..139H}%
  \BibitemOpen
  \bibfield  {author} {\bibinfo {author} {\bibfnamefont {K.}~\bibnamefont
  {{Holley-Bockelmann}}}\ and\ \bibinfo {author} {\bibfnamefont {F.~M.}\
  \bibnamefont {{Khan}}},\ }\href {\doibase 10.1088/0004-637X/810/2/139}
  {\bibfield  {journal} {\bibinfo  {journal} {\apj}\ }\textbf {\bibinfo
  {volume} {810}},\ \bibinfo {eid} {139} (\bibinfo {year} {2015})},\ \Eprint
  {http://arxiv.org/abs/1505.06203} {arXiv:1505.06203 [astro-ph.GA]}
  \BibitemShut {NoStop}%
\bibitem [{\citenamefont {{Vasiliev}}\ \emph {et~al.}(2015)\citenamefont
  {{Vasiliev}}, \citenamefont {{Antonini}},\ and\ \citenamefont
  {{Merritt}}}]{2015ApJ...810...49V}%
  \BibitemOpen
  \bibfield  {author} {\bibinfo {author} {\bibfnamefont {E.}~\bibnamefont
  {{Vasiliev}}}, \bibinfo {author} {\bibfnamefont {F.}~\bibnamefont
  {{Antonini}}}, \ and\ \bibinfo {author} {\bibfnamefont {D.}~\bibnamefont
  {{Merritt}}},\ }\href {\doibase 10.1088/0004-637X/810/1/49} {\bibfield
  {journal} {\bibinfo  {journal} {\apj}\ }\textbf {\bibinfo {volume} {810}},\
  \bibinfo {eid} {49} (\bibinfo {year} {2015})},\ \Eprint
  {http://arxiv.org/abs/1505.05480} {arXiv:1505.05480 [astro-ph.GA]}
  \BibitemShut {NoStop}%
\bibitem [{\citenamefont {{Gualandris}}\ \emph {et~al.}(2017)\citenamefont
  {{Gualandris}}, \citenamefont {{Read}}, \citenamefont {{Dehnen}},\ and\
  \citenamefont {{Bortolas}}}]{2017MNRAS.464.2301G}%
  \BibitemOpen
  \bibfield  {author} {\bibinfo {author} {\bibfnamefont {A.}~\bibnamefont
  {{Gualandris}}}, \bibinfo {author} {\bibfnamefont {J.~I.}\ \bibnamefont
  {{Read}}}, \bibinfo {author} {\bibfnamefont {W.}~\bibnamefont {{Dehnen}}}, \
  and\ \bibinfo {author} {\bibfnamefont {E.}~\bibnamefont {{Bortolas}}},\
  }\href {\doibase 10.1093/mnras/stw2528} {\bibfield  {journal} {\bibinfo
  {journal} {Mon. Not. Roy. Astron. Soc.}\ }\textbf {\bibinfo {volume} {464}},\
  \bibinfo {pages} {2301} (\bibinfo {year} {2017})},\ \Eprint
  {http://arxiv.org/abs/1609.09383} {arXiv:1609.09383 [astro-ph.GA]}
  \BibitemShut {NoStop}%
\bibitem [{\citenamefont {Escala}\ \emph {et~al.}(2005)\citenamefont {Escala},
  \citenamefont {Larson}, \citenamefont {Coppi},\ and\ \citenamefont
  {Mardones}}]{Escala:2004jh}%
  \BibitemOpen
  \bibfield  {author} {\bibinfo {author} {\bibfnamefont {A.}~\bibnamefont
  {Escala}}, \bibinfo {author} {\bibfnamefont {R.~B.}\ \bibnamefont {Larson}},
  \bibinfo {author} {\bibfnamefont {P.~S.}\ \bibnamefont {Coppi}}, \ and\
  \bibinfo {author} {\bibfnamefont {D.}~\bibnamefont {Mardones}},\ }\href
  {\doibase 10.1086/431747} {\bibfield  {journal} {\bibinfo  {journal}
  {Astrophys. J.}\ }\textbf {\bibinfo {volume} {630}},\ \bibinfo {pages} {152}
  (\bibinfo {year} {2005})},\ \Eprint {http://arxiv.org/abs/astro-ph/0406304}
  {arXiv:astro-ph/0406304} \BibitemShut {NoStop}%
\bibitem [{\citenamefont {Perets}\ \emph {et~al.}(2007)\citenamefont {Perets},
  \citenamefont {Hopman},\ and\ \citenamefont {Alexander}}]{Perets:2006bz}%
  \BibitemOpen
  \bibfield  {author} {\bibinfo {author} {\bibfnamefont {H.~B.}\ \bibnamefont
  {Perets}}, \bibinfo {author} {\bibfnamefont {C.}~\bibnamefont {Hopman}}, \
  and\ \bibinfo {author} {\bibfnamefont {T.}~\bibnamefont {Alexander}},\ }\href
  {\doibase 10.1086/510377} {\bibfield  {journal} {\bibinfo  {journal}
  {Astrophys. J.}\ }\textbf {\bibinfo {volume} {656}},\ \bibinfo {pages} {709}
  (\bibinfo {year} {2007})},\ \Eprint {http://arxiv.org/abs/astro-ph/0606443}
  {arXiv:astro-ph/0606443} \BibitemShut {NoStop}%
\bibitem [{\citenamefont {Perets}\ and\ \citenamefont
  {Alexander}(2008)}]{Perets:2007nc}%
  \BibitemOpen
  \bibfield  {author} {\bibinfo {author} {\bibfnamefont {H.~B.}\ \bibnamefont
  {Perets}}\ and\ \bibinfo {author} {\bibfnamefont {T.}~\bibnamefont
  {Alexander}},\ }\href {\doibase 10.1086/527525} {\bibfield  {journal}
  {\bibinfo  {journal} {Astrophys. J.}\ }\textbf {\bibinfo {volume} {677}},\
  \bibinfo {pages} {146} (\bibinfo {year} {2008})},\ \Eprint
  {http://arxiv.org/abs/0705.2123} {arXiv:0705.2123 [astro-ph]} \BibitemShut
  {NoStop}%
\bibitem [{\citenamefont {Rindler-Daller}\ and\ \citenamefont
  {Shapiro}(2012)}]{Rindler-Daller:2011afd}%
  \BibitemOpen
  \bibfield  {author} {\bibinfo {author} {\bibfnamefont {T.}~\bibnamefont
  {Rindler-Daller}}\ and\ \bibinfo {author} {\bibfnamefont {P.~R.}\
  \bibnamefont {Shapiro}},\ }\href {\doibase 10.1111/j.1365-2966.2012.20588.x}
  {\bibfield  {journal} {\bibinfo  {journal} {Mon. Not. Roy. Astron. Soc.}\
  }\textbf {\bibinfo {volume} {422}},\ \bibinfo {pages} {135} (\bibinfo {year}
  {2012})},\ \Eprint {http://arxiv.org/abs/1106.1256} {arXiv:1106.1256
  [astro-ph.CO]} \BibitemShut {NoStop}%
\bibitem [{\citenamefont {Berezhiani}\ and\ \citenamefont
  {Khoury}(2016)}]{Berezhiani:2015pia}%
  \BibitemOpen
  \bibfield  {author} {\bibinfo {author} {\bibfnamefont {L.}~\bibnamefont
  {Berezhiani}}\ and\ \bibinfo {author} {\bibfnamefont {J.}~\bibnamefont
  {Khoury}},\ }\href {\doibase 10.1016/j.physletb.2015.12.054} {\bibfield
  {journal} {\bibinfo  {journal} {Phys. Lett. B}\ }\textbf {\bibinfo {volume}
  {753}},\ \bibinfo {pages} {639} (\bibinfo {year} {2016})},\ \Eprint
  {http://arxiv.org/abs/1506.07877} {arXiv:1506.07877 [astro-ph.CO]}
  \BibitemShut {NoStop}%
\bibitem [{\citenamefont {Armengaud}\ \emph {et~al.}(2017)\citenamefont
  {Armengaud}, \citenamefont {Palanque-Delabrouille}, \citenamefont {Y\`eche},
  \citenamefont {Marsh},\ and\ \citenamefont {Baur}}]{Armengaud:2017nkf}%
  \BibitemOpen
  \bibfield  {author} {\bibinfo {author} {\bibfnamefont {E.}~\bibnamefont
  {Armengaud}}, \bibinfo {author} {\bibfnamefont {N.}~\bibnamefont
  {Palanque-Delabrouille}}, \bibinfo {author} {\bibfnamefont {C.}~\bibnamefont
  {Y\`eche}}, \bibinfo {author} {\bibfnamefont {D.~J.~E.}\ \bibnamefont
  {Marsh}}, \ and\ \bibinfo {author} {\bibfnamefont {J.}~\bibnamefont {Baur}},\
  }\href {\doibase 10.1093/mnras/stx1870} {\bibfield  {journal} {\bibinfo
  {journal} {Mon. Not. Roy. Astron. Soc.}\ }\textbf {\bibinfo {volume} {471}},\
  \bibinfo {pages} {4606} (\bibinfo {year} {2017})},\ \Eprint
  {http://arxiv.org/abs/1703.09126} {arXiv:1703.09126 [astro-ph.CO]}
  \BibitemShut {NoStop}%
\bibitem [{\citenamefont {Ir\v{s}i\v{c}}\ \emph {et~al.}(2017)\citenamefont
  {Ir\v{s}i\v{c}}, \citenamefont {Viel}, \citenamefont {Haehnelt},
  \citenamefont {Bolton},\ and\ \citenamefont {Becker}}]{Irsic:2017yje}%
  \BibitemOpen
  \bibfield  {author} {\bibinfo {author} {\bibfnamefont {V.}~\bibnamefont
  {Ir\v{s}i\v{c}}}, \bibinfo {author} {\bibfnamefont {M.}~\bibnamefont {Viel}},
  \bibinfo {author} {\bibfnamefont {M.~G.}\ \bibnamefont {Haehnelt}}, \bibinfo
  {author} {\bibfnamefont {J.~S.}\ \bibnamefont {Bolton}}, \ and\ \bibinfo
  {author} {\bibfnamefont {G.~D.}\ \bibnamefont {Becker}},\ }\href {\doibase
  10.1103/PhysRevLett.119.031302} {\bibfield  {journal} {\bibinfo  {journal}
  {Phys. Rev. Lett.}\ }\textbf {\bibinfo {volume} {119}},\ \bibinfo {pages}
  {031302} (\bibinfo {year} {2017})},\ \Eprint
  {http://arxiv.org/abs/1703.04683} {arXiv:1703.04683 [astro-ph.CO]}
  \BibitemShut {NoStop}%
\bibitem [{\citenamefont {Nori}\ \emph {et~al.}(2019)\citenamefont {Nori},
  \citenamefont {Murgia}, \citenamefont {Ir\v{s}i\v{c}}, \citenamefont
  {Baldi},\ and\ \citenamefont {Viel}}]{Nori:2018pka}%
  \BibitemOpen
  \bibfield  {author} {\bibinfo {author} {\bibfnamefont {M.}~\bibnamefont
  {Nori}}, \bibinfo {author} {\bibfnamefont {R.}~\bibnamefont {Murgia}},
  \bibinfo {author} {\bibfnamefont {V.}~\bibnamefont {Ir\v{s}i\v{c}}}, \bibinfo
  {author} {\bibfnamefont {M.}~\bibnamefont {Baldi}}, \ and\ \bibinfo {author}
  {\bibfnamefont {M.}~\bibnamefont {Viel}},\ }\href {\doibase
  10.1093/mnras/sty2888} {\bibfield  {journal} {\bibinfo  {journal} {Mon. Not.
  Roy. Astron. Soc.}\ }\textbf {\bibinfo {volume} {482}},\ \bibinfo {pages}
  {3227} (\bibinfo {year} {2019})},\ \Eprint {http://arxiv.org/abs/1809.09619}
  {arXiv:1809.09619 [astro-ph.CO]} \BibitemShut {NoStop}%
\bibitem [{\citenamefont {Rogers}\ and\ \citenamefont
  {Peiris}(2021)}]{Rogers:2020ltq}%
  \BibitemOpen
  \bibfield  {author} {\bibinfo {author} {\bibfnamefont {K.~K.}\ \bibnamefont
  {Rogers}}\ and\ \bibinfo {author} {\bibfnamefont {H.~V.}\ \bibnamefont
  {Peiris}},\ }\href {\doibase 10.1103/PhysRevLett.126.071302} {\bibfield
  {journal} {\bibinfo  {journal} {Phys. Rev. Lett.}\ }\textbf {\bibinfo
  {volume} {126}},\ \bibinfo {pages} {071302} (\bibinfo {year} {2021})},\
  \Eprint {http://arxiv.org/abs/2007.12705} {arXiv:2007.12705 [astro-ph.CO]}
  \BibitemShut {NoStop}%
\bibitem [{\citenamefont {Marsh}\ and\ \citenamefont
  {Niemeyer}(2019)}]{Marsh:2018zyw}%
  \BibitemOpen
  \bibfield  {author} {\bibinfo {author} {\bibfnamefont {D.~J.~E.}\
  \bibnamefont {Marsh}}\ and\ \bibinfo {author} {\bibfnamefont {J.~C.}\
  \bibnamefont {Niemeyer}},\ }\href {\doibase 10.1103/PhysRevLett.123.051103}
  {\bibfield  {journal} {\bibinfo  {journal} {Phys. Rev. Lett.}\ }\textbf
  {\bibinfo {volume} {123}},\ \bibinfo {pages} {051103} (\bibinfo {year}
  {2019})},\ \Eprint {http://arxiv.org/abs/1810.08543} {arXiv:1810.08543
  [astro-ph.CO]} \BibitemShut {NoStop}%
\bibitem [{\citenamefont {Dalal}\ and\ \citenamefont
  {Kravtsov}(2022)}]{Dalal:2022rmp}%
  \BibitemOpen
  \bibfield  {author} {\bibinfo {author} {\bibfnamefont {N.}~\bibnamefont
  {Dalal}}\ and\ \bibinfo {author} {\bibfnamefont {A.}~\bibnamefont
  {Kravtsov}},\ }\href {\doibase 10.1103/PhysRevD.106.063517} {\bibfield
  {journal} {\bibinfo  {journal} {Phys. Rev. D}\ }\textbf {\bibinfo {volume}
  {106}},\ \bibinfo {pages} {063517} (\bibinfo {year} {2022})},\ \Eprint
  {http://arxiv.org/abs/2203.05750} {arXiv:2203.05750 [astro-ph.CO]}
  \BibitemShut {NoStop}%
\bibitem [{\citenamefont {{Amin}}\ and\ \citenamefont
  {{Mirbabayi}}(2022)}]{Amin:2022nlh}%
  \BibitemOpen
  \bibfield  {author} {\bibinfo {author} {\bibfnamefont {M.~A.}\ \bibnamefont
  {{Amin}}}\ and\ \bibinfo {author} {\bibfnamefont {M.}~\bibnamefont
  {{Mirbabayi}}},\ }\href {\doibase 10.48550/arXiv.2211.09775} {\bibfield
  {journal} {\bibinfo  {journal} {arXiv e-prints}\ ,\ \bibinfo {eid}
  {arXiv:2211.09775}} (\bibinfo {year} {2022})},\ \Eprint
  {http://arxiv.org/abs/2211.09775} {arXiv:2211.09775 [hep-ph]} \BibitemShut
  {NoStop}%
\bibitem [{\citenamefont {Hui}(2021)}]{Hui:2021tkt}%
  \BibitemOpen
  \bibfield  {author} {\bibinfo {author} {\bibfnamefont {L.}~\bibnamefont
  {Hui}},\ }\href {\doibase 10.1146/annurev-astro-120920-010024} {\bibfield
  {journal} {\bibinfo  {journal} {Ann. Rev. Astron. Astrophys.}\ }\textbf
  {\bibinfo {volume} {59}},\ \bibinfo {pages} {247} (\bibinfo {year} {2021})},\
  \Eprint {http://arxiv.org/abs/2101.11735} {arXiv:2101.11735 [astro-ph.CO]}
  \BibitemShut {NoStop}%
\bibitem [{\citenamefont {Ferreira}(2021)}]{Ferreira:2020fam}%
  \BibitemOpen
  \bibfield  {author} {\bibinfo {author} {\bibfnamefont {E.~G.~M.}\
  \bibnamefont {Ferreira}},\ }\href {\doibase 10.1007/s00159-021-00135-6}
  {\bibfield  {journal} {\bibinfo  {journal} {Astron. Astrophys. Rev.}\
  }\textbf {\bibinfo {volume} {29}},\ \bibinfo {pages} {7} (\bibinfo {year}
  {2021})},\ \Eprint {http://arxiv.org/abs/2005.03254} {arXiv:2005.03254
  [astro-ph.CO]} \BibitemShut {NoStop}%
\bibitem [{\citenamefont {Schive}\ \emph
  {et~al.}(2014{\natexlab{a}})\citenamefont {Schive}, \citenamefont {Chiueh},\
  and\ \citenamefont {Broadhurst}}]{Schive:2014dra}%
  \BibitemOpen
  \bibfield  {author} {\bibinfo {author} {\bibfnamefont {H.-Y.}\ \bibnamefont
  {Schive}}, \bibinfo {author} {\bibfnamefont {T.}~\bibnamefont {Chiueh}}, \
  and\ \bibinfo {author} {\bibfnamefont {T.}~\bibnamefont {Broadhurst}},\
  }\href {\doibase 10.1038/nphys2996} {\bibfield  {journal} {\bibinfo
  {journal} {Nature Phys.}\ }\textbf {\bibinfo {volume} {10}},\ \bibinfo
  {pages} {496} (\bibinfo {year} {2014}{\natexlab{a}})},\ \Eprint
  {http://arxiv.org/abs/1406.6586} {arXiv:1406.6586 [astro-ph.GA]} \BibitemShut
  {NoStop}%
\bibitem [{\citenamefont {Hui}\ \emph {et~al.}(2017)\citenamefont {Hui},
  \citenamefont {Ostriker}, \citenamefont {Tremaine},\ and\ \citenamefont
  {Witten}}]{Hui:2016ltb}%
  \BibitemOpen
  \bibfield  {author} {\bibinfo {author} {\bibfnamefont {L.}~\bibnamefont
  {Hui}}, \bibinfo {author} {\bibfnamefont {J.~P.}\ \bibnamefont {Ostriker}},
  \bibinfo {author} {\bibfnamefont {S.}~\bibnamefont {Tremaine}}, \ and\
  \bibinfo {author} {\bibfnamefont {E.}~\bibnamefont {Witten}},\ }\href
  {\doibase 10.1103/PhysRevD.95.043541} {\bibfield  {journal} {\bibinfo
  {journal} {Phys. Rev. D}\ }\textbf {\bibinfo {volume} {95}},\ \bibinfo
  {pages} {043541} (\bibinfo {year} {2017})},\ \Eprint
  {http://arxiv.org/abs/1610.08297} {arXiv:1610.08297 [astro-ph.CO]}
  \BibitemShut {NoStop}%
\bibitem [{\citenamefont {Zupancic}\ and\ \citenamefont
  {Widrow}(2023)}]{Zupancic:2023qgj}%
  \BibitemOpen
  \bibfield  {author} {\bibinfo {author} {\bibfnamefont {B.}~\bibnamefont
  {Zupancic}}\ and\ \bibinfo {author} {\bibfnamefont {L.~M.}\ \bibnamefont
  {Widrow}},\ }\href@noop {} {\bibfield  {journal} {\bibinfo  {journal} {arXiv
  e-print}\ } (\bibinfo {year} {2023})},\ \Eprint
  {http://arxiv.org/abs/2311.13352} {arXiv:2311.13352 [astro-ph.CO]}
  \BibitemShut {NoStop}%
\bibitem [{\citenamefont {Veltmaat}\ \emph {et~al.}(2018)\citenamefont
  {Veltmaat}, \citenamefont {Niemeyer},\ and\ \citenamefont
  {Schwabe}}]{Veltmaat:2018dfz}%
  \BibitemOpen
  \bibfield  {author} {\bibinfo {author} {\bibfnamefont {J.}~\bibnamefont
  {Veltmaat}}, \bibinfo {author} {\bibfnamefont {J.~C.}\ \bibnamefont
  {Niemeyer}}, \ and\ \bibinfo {author} {\bibfnamefont {B.}~\bibnamefont
  {Schwabe}},\ }\href {\doibase 10.1103/PhysRevD.98.043509} {\bibfield
  {journal} {\bibinfo  {journal} {Phys. Rev. D}\ }\textbf {\bibinfo {volume}
  {98}},\ \bibinfo {pages} {043509} (\bibinfo {year} {2018})},\ \Eprint
  {http://arxiv.org/abs/1804.09647} {arXiv:1804.09647 [astro-ph.CO]}
  \BibitemShut {NoStop}%
\bibitem [{\citenamefont {Begelman}\ \emph {et~al.}(1980)\citenamefont
  {Begelman}, \citenamefont {Blandford},\ and\ \citenamefont
  {Rees}}]{Begelman:1980vb}%
  \BibitemOpen
  \bibfield  {author} {\bibinfo {author} {\bibfnamefont {M.~C.}\ \bibnamefont
  {Begelman}}, \bibinfo {author} {\bibfnamefont {R.~D.}\ \bibnamefont
  {Blandford}}, \ and\ \bibinfo {author} {\bibfnamefont {M.~J.}\ \bibnamefont
  {Rees}},\ }\href {\doibase 10.1038/287307a0} {\bibfield  {journal} {\bibinfo
  {journal} {Nature}\ }\textbf {\bibinfo {volume} {287}},\ \bibinfo {pages}
  {307} (\bibinfo {year} {1980})}\BibitemShut {NoStop}%
\bibitem [{\citenamefont {{Binney}}\ and\ \citenamefont
  {{Tremaine}}(2008)}]{2008gady.book.....B}%
  \BibitemOpen
  \bibfield  {author} {\bibinfo {author} {\bibfnamefont {J.}~\bibnamefont
  {{Binney}}}\ and\ \bibinfo {author} {\bibfnamefont {S.}~\bibnamefont
  {{Tremaine}}},\ }\href@noop {} {\emph {\bibinfo {title} {{Galactic Dynamics:
  Second Edition}}}}\ (\bibinfo  {publisher} {Princeton University Press},\
  \bibinfo {year} {2008})\BibitemShut {NoStop}%
\bibitem [{\citenamefont {Chandrasekhar}(1943)}]{Chandrasekhar:1943ys}%
  \BibitemOpen
  \bibfield  {author} {\bibinfo {author} {\bibfnamefont {S.}~\bibnamefont
  {Chandrasekhar}},\ }\href {\doibase 10.1086/144517} {\bibfield  {journal}
  {\bibinfo  {journal} {Astrophys. J.}\ }\textbf {\bibinfo {volume} {97}},\
  \bibinfo {pages} {255} (\bibinfo {year} {1943})}\BibitemShut {NoStop}%
\bibitem [{\citenamefont {{Aceves}}\ and\ \citenamefont
  {{Colosimo}}(2007)}]{2007AmJPh..75..139A}%
  \BibitemOpen
  \bibfield  {author} {\bibinfo {author} {\bibfnamefont {H.}~\bibnamefont
  {{Aceves}}}\ and\ \bibinfo {author} {\bibfnamefont {M.}~\bibnamefont
  {{Colosimo}}},\ }\href {\doibase 10.1119/1.2388968} {\bibfield  {journal}
  {\bibinfo  {journal} {American Journal of Physics}\ }\textbf {\bibinfo
  {volume} {75}},\ \bibinfo {pages} {139} (\bibinfo {year} {2007})},\ \Eprint
  {http://arxiv.org/abs/physics/0603066} {arXiv:physics/0603066
  [physics.class-ph]} \BibitemShut {NoStop}%
\bibitem [{\citenamefont {{Hills}}\ and\ \citenamefont
  {{Fullerton}}(1980)}]{1980AJ.....85.1281H}%
  \BibitemOpen
  \bibfield  {author} {\bibinfo {author} {\bibfnamefont {J.~G.}\ \bibnamefont
  {{Hills}}}\ and\ \bibinfo {author} {\bibfnamefont {L.~W.}\ \bibnamefont
  {{Fullerton}}},\ }\href {\doibase 10.1086/112798} {\bibfield  {journal}
  {\bibinfo  {journal} {AJ}\ }\textbf {\bibinfo {volume} {85}},\ \bibinfo
  {pages} {1281} (\bibinfo {year} {1980})}\BibitemShut {NoStop}%
\bibitem [{\citenamefont {Quinlan}(1996)}]{Quinlan:1996vp}%
  \BibitemOpen
  \bibfield  {author} {\bibinfo {author} {\bibfnamefont {G.~D.}\ \bibnamefont
  {Quinlan}},\ }\href {\doibase 10.1016/S1384-1076(96)00003-6} {\bibfield
  {journal} {\bibinfo  {journal} {New Astron.}\ }\textbf {\bibinfo {volume}
  {1}},\ \bibinfo {pages} {35} (\bibinfo {year} {1996})},\ \Eprint
  {http://arxiv.org/abs/astro-ph/9601092} {arXiv:astro-ph/9601092} \BibitemShut
  {NoStop}%
\bibitem [{\citenamefont {{Colpi}}(2014)}]{2014SSRv..183..189C}%
  \BibitemOpen
  \bibfield  {author} {\bibinfo {author} {\bibfnamefont {M.}~\bibnamefont
  {{Colpi}}},\ }\href {\doibase 10.1007/s11214-014-0067-1} {\bibfield
  {journal} {\bibinfo  {journal} {Space Science Reviews}\ }\textbf {\bibinfo
  {volume} {183}},\ \bibinfo {pages} {189} (\bibinfo {year} {2014})},\ \Eprint
  {http://arxiv.org/abs/1407.3102} {arXiv:1407.3102 [astro-ph.GA]} \BibitemShut
  {NoStop}%
\bibitem [{\citenamefont {Celoria}\ \emph {et~al.}(2018)\citenamefont
  {Celoria}, \citenamefont {Oliveri}, \citenamefont {Sesana},\ and\
  \citenamefont {Mapelli}}]{Celoria:2018mzr}%
  \BibitemOpen
  \bibfield  {author} {\bibinfo {author} {\bibfnamefont {M.}~\bibnamefont
  {Celoria}}, \bibinfo {author} {\bibfnamefont {R.}~\bibnamefont {Oliveri}},
  \bibinfo {author} {\bibfnamefont {A.}~\bibnamefont {Sesana}}, \ and\ \bibinfo
  {author} {\bibfnamefont {M.}~\bibnamefont {Mapelli}}\ }(\bibinfo {year}
  {2018})\ \Eprint {http://arxiv.org/abs/1807.11489} {arXiv:1807.11489
  [astro-ph.GA]} \BibitemShut {NoStop}%
\bibitem [{\citenamefont {{Merritt}}(2013)}]{2013CQGra..30x4005M}%
  \BibitemOpen
  \bibfield  {author} {\bibinfo {author} {\bibfnamefont {D.}~\bibnamefont
  {{Merritt}}},\ }\href {\doibase 10.1088/0264-9381/30/24/244005} {\bibfield
  {journal} {\bibinfo  {journal} {Classical and Quantum Gravity}\ }\textbf
  {\bibinfo {volume} {30}},\ \bibinfo {eid} {244005} (\bibinfo {year}
  {2013})},\ \Eprint {http://arxiv.org/abs/1307.3268} {arXiv:1307.3268
  [astro-ph.GA]} \BibitemShut {NoStop}%
\bibitem [{\citenamefont {Peters}(1964)}]{Peters:1964zz}%
  \BibitemOpen
  \bibfield  {author} {\bibinfo {author} {\bibfnamefont {P.~C.}\ \bibnamefont
  {Peters}},\ }\href {\doibase 10.1103/PhysRev.136.B1224} {\bibfield  {journal}
  {\bibinfo  {journal} {Phys. Rev.}\ }\textbf {\bibinfo {volume} {136}},\
  \bibinfo {pages} {B1224} (\bibinfo {year} {1964})}\BibitemShut {NoStop}%
\bibitem [{\citenamefont {Merritt}\ and\ \citenamefont
  {Milosavljevic}(2005)}]{Merritt:2004gc}%
  \BibitemOpen
  \bibfield  {author} {\bibinfo {author} {\bibfnamefont {D.}~\bibnamefont
  {Merritt}}\ and\ \bibinfo {author} {\bibfnamefont {M.}~\bibnamefont
  {Milosavljevic}},\ }\href {\doibase 10.12942/lrr-2005-8} {\bibfield
  {journal} {\bibinfo  {journal} {Living Rev. Rel.}\ }\textbf {\bibinfo
  {volume} {8}},\ \bibinfo {pages} {8} (\bibinfo {year} {2005})},\ \Eprint
  {http://arxiv.org/abs/astro-ph/0410364} {arXiv:astro-ph/0410364} \BibitemShut
  {NoStop}%
\bibitem [{\citenamefont {Ruffini}\ and\ \citenamefont
  {Bonazzola}(1969)}]{Ruffini:1969qy}%
  \BibitemOpen
  \bibfield  {author} {\bibinfo {author} {\bibfnamefont {R.}~\bibnamefont
  {Ruffini}}\ and\ \bibinfo {author} {\bibfnamefont {S.}~\bibnamefont
  {Bonazzola}},\ }\href {\doibase 10.1103/PhysRev.187.1767} {\bibfield
  {journal} {\bibinfo  {journal} {Phys. Rev.}\ }\textbf {\bibinfo {volume}
  {187}},\ \bibinfo {pages} {1767} (\bibinfo {year} {1969})}\BibitemShut
  {NoStop}%
\bibitem [{\citenamefont {Schive}\ \emph
  {et~al.}(2014{\natexlab{b}})\citenamefont {Schive}, \citenamefont {Liao},
  \citenamefont {Woo}, \citenamefont {Wong}, \citenamefont {Chiueh},
  \citenamefont {Broadhurst},\ and\ \citenamefont {Hwang}}]{Schive:2014hza}%
  \BibitemOpen
  \bibfield  {author} {\bibinfo {author} {\bibfnamefont {H.-Y.}\ \bibnamefont
  {Schive}}, \bibinfo {author} {\bibfnamefont {M.-H.}\ \bibnamefont {Liao}},
  \bibinfo {author} {\bibfnamefont {T.-P.}\ \bibnamefont {Woo}}, \bibinfo
  {author} {\bibfnamefont {S.-K.}\ \bibnamefont {Wong}}, \bibinfo {author}
  {\bibfnamefont {T.}~\bibnamefont {Chiueh}}, \bibinfo {author} {\bibfnamefont
  {T.}~\bibnamefont {Broadhurst}}, \ and\ \bibinfo {author} {\bibfnamefont
  {W.~Y.~P.}\ \bibnamefont {Hwang}},\ }\href {\doibase
  10.1103/PhysRevLett.113.261302} {\bibfield  {journal} {\bibinfo  {journal}
  {Phys. Rev. Lett.}\ }\textbf {\bibinfo {volume} {113}},\ \bibinfo {pages}
  {261302} (\bibinfo {year} {2014}{\natexlab{b}})},\ \Eprint
  {http://arxiv.org/abs/1407.7762} {arXiv:1407.7762 [astro-ph.GA]} \BibitemShut
  {NoStop}%
\bibitem [{\citenamefont {Bar-Or}\ \emph {et~al.}(2019)\citenamefont {Bar-Or},
  \citenamefont {Fouvry},\ and\ \citenamefont {Tremaine}}]{Bar-Or:2018pxz}%
  \BibitemOpen
  \bibfield  {author} {\bibinfo {author} {\bibfnamefont {B.}~\bibnamefont
  {Bar-Or}}, \bibinfo {author} {\bibfnamefont {J.-B.}\ \bibnamefont {Fouvry}},
  \ and\ \bibinfo {author} {\bibfnamefont {S.}~\bibnamefont {Tremaine}},\
  }\href {\doibase 10.3847/1538-4357/aaf28c} {\bibfield  {journal} {\bibinfo
  {journal} {Astrophys. J.}\ }\textbf {\bibinfo {volume} {871}},\ \bibinfo
  {pages} {28} (\bibinfo {year} {2019})},\ \Eprint
  {http://arxiv.org/abs/1809.07673} {arXiv:1809.07673 [astro-ph.GA]}
  \BibitemShut {NoStop}%
\bibitem [{\citenamefont {Lancaster}\ \emph {et~al.}(2020)\citenamefont
  {Lancaster}, \citenamefont {Giovanetti}, \citenamefont {Mocz}, \citenamefont
  {Kahn}, \citenamefont {Lisanti},\ and\ \citenamefont
  {Spergel}}]{Lancaster:2019mde}%
  \BibitemOpen
  \bibfield  {author} {\bibinfo {author} {\bibfnamefont {L.}~\bibnamefont
  {Lancaster}}, \bibinfo {author} {\bibfnamefont {C.}~\bibnamefont
  {Giovanetti}}, \bibinfo {author} {\bibfnamefont {P.}~\bibnamefont {Mocz}},
  \bibinfo {author} {\bibfnamefont {Y.}~\bibnamefont {Kahn}}, \bibinfo {author}
  {\bibfnamefont {M.}~\bibnamefont {Lisanti}}, \ and\ \bibinfo {author}
  {\bibfnamefont {D.~N.}\ \bibnamefont {Spergel}},\ }\href {\doibase
  10.1088/1475-7516/2020/01/001} {\bibfield  {journal} {\bibinfo  {journal}
  {JCAP}\ }\textbf {\bibinfo {volume} {01}},\ \bibinfo {pages} {001} (\bibinfo
  {year} {2020})},\ \Eprint {http://arxiv.org/abs/1909.06381} {arXiv:1909.06381
  [astro-ph.CO]} \BibitemShut {NoStop}%
\bibitem [{\citenamefont {{Bar-Or}}\ \emph {et~al.}(2021)\citenamefont
  {{Bar-Or}}, \citenamefont {{Fouvry}},\ and\ \citenamefont
  {{Tremaine}}}]{2021ApJ...915...27B}%
  \BibitemOpen
  \bibfield  {author} {\bibinfo {author} {\bibfnamefont {B.}~\bibnamefont
  {{Bar-Or}}}, \bibinfo {author} {\bibfnamefont {J.-B.}\ \bibnamefont
  {{Fouvry}}}, \ and\ \bibinfo {author} {\bibfnamefont {S.}~\bibnamefont
  {{Tremaine}}},\ }\href {\doibase 10.3847/1538-4357/abfb66} {\bibfield
  {journal} {\bibinfo  {journal} {\apj}\ }\textbf {\bibinfo {volume} {915}},\
  \bibinfo {eid} {27} (\bibinfo {year} {2021})},\ \Eprint
  {http://arxiv.org/abs/2010.10212} {arXiv:2010.10212 [astro-ph.GA]}
  \BibitemShut {NoStop}%
\bibitem [{\citenamefont {Seidel}\ and\ \citenamefont
  {Suen}(1994)}]{Seidel:1993zk}%
  \BibitemOpen
  \bibfield  {author} {\bibinfo {author} {\bibfnamefont {E.}~\bibnamefont
  {Seidel}}\ and\ \bibinfo {author} {\bibfnamefont {W.-M.}\ \bibnamefont
  {Suen}},\ }\href {\doibase 10.1103/PhysRevLett.72.2516} {\bibfield  {journal}
  {\bibinfo  {journal} {Phys. Rev. Lett.}\ }\textbf {\bibinfo {volume} {72}},\
  \bibinfo {pages} {2516} (\bibinfo {year} {1994})},\ \Eprint
  {http://arxiv.org/abs/gr-qc/9309015} {arXiv:gr-qc/9309015} \BibitemShut
  {NoStop}%
\bibitem [{\citenamefont {Guzman}\ and\ \citenamefont
  {Urena-Lopez}(2006)}]{Guzman:2006yc}%
  \BibitemOpen
  \bibfield  {author} {\bibinfo {author} {\bibfnamefont {F.~S.}\ \bibnamefont
  {Guzman}}\ and\ \bibinfo {author} {\bibfnamefont {L.~A.}\ \bibnamefont
  {Urena-Lopez}},\ }\href {\doibase 10.1086/504508} {\bibfield  {journal}
  {\bibinfo  {journal} {Astrophys. J.}\ }\textbf {\bibinfo {volume} {645}},\
  \bibinfo {pages} {814} (\bibinfo {year} {2006})},\ \Eprint
  {http://arxiv.org/abs/astro-ph/0603613} {arXiv:astro-ph/0603613} \BibitemShut
  {NoStop}%
\bibitem [{\citenamefont {Schwabe}\ \emph {et~al.}(2016)\citenamefont
  {Schwabe}, \citenamefont {Niemeyer},\ and\ \citenamefont
  {Engels}}]{Schwabe:2016rze}%
  \BibitemOpen
  \bibfield  {author} {\bibinfo {author} {\bibfnamefont {B.}~\bibnamefont
  {Schwabe}}, \bibinfo {author} {\bibfnamefont {J.~C.}\ \bibnamefont
  {Niemeyer}}, \ and\ \bibinfo {author} {\bibfnamefont {J.~F.}\ \bibnamefont
  {Engels}},\ }\href {\doibase 10.1103/PhysRevD.94.043513} {\bibfield
  {journal} {\bibinfo  {journal} {Phys. Rev. D}\ }\textbf {\bibinfo {volume}
  {94}},\ \bibinfo {pages} {043513} (\bibinfo {year} {2016})},\ \Eprint
  {http://arxiv.org/abs/1606.05151} {arXiv:1606.05151 [astro-ph.CO]}
  \BibitemShut {NoStop}%
\bibitem [{\citenamefont {Alvarez-R\'\i{}os}\ \emph {et~al.}(2023)\citenamefont
  {Alvarez-R\'\i{}os}, \citenamefont {Guzm\'an},\ and\ \citenamefont
  {Shapiro}}]{Alvarez-Rios:2023cch}%
  \BibitemOpen
  \bibfield  {author} {\bibinfo {author} {\bibfnamefont {I.}~\bibnamefont
  {Alvarez-R\'\i{}os}}, \bibinfo {author} {\bibfnamefont {F.~S.}\ \bibnamefont
  {Guzm\'an}}, \ and\ \bibinfo {author} {\bibfnamefont {P.~R.}\ \bibnamefont
  {Shapiro}},\ }\href {\doibase 10.1103/PhysRevD.107.123524} {\bibfield
  {journal} {\bibinfo  {journal} {Phys. Rev. D}\ }\textbf {\bibinfo {volume}
  {107}},\ \bibinfo {pages} {123524} (\bibinfo {year} {2023})},\ \Eprint
  {http://arxiv.org/abs/2304.03419} {arXiv:2304.03419 [astro-ph.CO]}
  \BibitemShut {NoStop}%
\bibitem [{\citenamefont {{Koo}}\ \emph {et~al.}(2023)\citenamefont {{Koo}},
  \citenamefont {{Bak}}, \citenamefont {{Park}}, \citenamefont {{Hong}},\ and\
  \citenamefont {{Lee}}}]{2023arXiv231103412K}%
  \BibitemOpen
  \bibfield  {author} {\bibinfo {author} {\bibfnamefont {H.}~\bibnamefont
  {{Koo}}}, \bibinfo {author} {\bibfnamefont {D.}~\bibnamefont {{Bak}}},
  \bibinfo {author} {\bibfnamefont {I.}~\bibnamefont {{Park}}}, \bibinfo
  {author} {\bibfnamefont {S.~E.}\ \bibnamefont {{Hong}}}, \ and\ \bibinfo
  {author} {\bibfnamefont {J.-W.}\ \bibnamefont {{Lee}}},\ }\href@noop {}
  {\bibfield  {journal} {\bibinfo  {journal} {arXiv e-prints}\ } (\bibinfo
  {year} {2023})},\ \Eprint {http://arxiv.org/abs/2311.03412} {arXiv:2311.03412
  [astro-ph.GA]} \BibitemShut {NoStop}%
\bibitem [{\citenamefont {Wang}\ and\ \citenamefont
  {Easther}(2022)}]{Wang:2021udl}%
  \BibitemOpen
  \bibfield  {author} {\bibinfo {author} {\bibfnamefont {Y.}~\bibnamefont
  {Wang}}\ and\ \bibinfo {author} {\bibfnamefont {R.}~\bibnamefont {Easther}},\
  }\href {\doibase 10.1103/PhysRevD.105.063523} {\bibfield  {journal} {\bibinfo
   {journal} {Phys. Rev. D}\ }\textbf {\bibinfo {volume} {105}},\ \bibinfo
  {pages} {063523} (\bibinfo {year} {2022})},\ \Eprint
  {http://arxiv.org/abs/2110.03428} {arXiv:2110.03428 [gr-qc]} \BibitemShut
  {NoStop}%
\bibitem [{\citenamefont {Cardoso}\ \emph {et~al.}(2022)\citenamefont
  {Cardoso}, \citenamefont {Ikeda}, \citenamefont {Vicente},\ and\
  \citenamefont {Zilh\~ao}}]{Cardoso:2022nzc}%
  \BibitemOpen
  \bibfield  {author} {\bibinfo {author} {\bibfnamefont {V.}~\bibnamefont
  {Cardoso}}, \bibinfo {author} {\bibfnamefont {T.}~\bibnamefont {Ikeda}},
  \bibinfo {author} {\bibfnamefont {R.}~\bibnamefont {Vicente}}, \ and\
  \bibinfo {author} {\bibfnamefont {M.}~\bibnamefont {Zilh\~ao}},\ }\href
  {\doibase 10.1103/PhysRevD.106.L121302} {\bibfield  {journal} {\bibinfo
  {journal} {Phys. Rev. D}\ }\textbf {\bibinfo {volume} {106}},\ \bibinfo
  {pages} {L121302} (\bibinfo {year} {2022})},\ \Eprint
  {http://arxiv.org/abs/2207.09469} {arXiv:2207.09469 [gr-qc]} \BibitemShut
  {NoStop}%
\bibitem [{\citenamefont {Navarro}\ \emph {et~al.}(1997)\citenamefont
  {Navarro}, \citenamefont {Frenk},\ and\ \citenamefont
  {White}}]{Navarro:1996gj}%
  \BibitemOpen
  \bibfield  {author} {\bibinfo {author} {\bibfnamefont {J.~F.}\ \bibnamefont
  {Navarro}}, \bibinfo {author} {\bibfnamefont {C.~S.}\ \bibnamefont {Frenk}},
  \ and\ \bibinfo {author} {\bibfnamefont {S.~D.~M.}\ \bibnamefont {White}},\
  }\href {\doibase 10.1086/304888} {\bibfield  {journal} {\bibinfo  {journal}
  {Astrophys. J.}\ }\textbf {\bibinfo {volume} {490}},\ \bibinfo {pages} {493}
  (\bibinfo {year} {1997})},\ \Eprint {http://arxiv.org/abs/astro-ph/9611107}
  {arXiv:astro-ph/9611107} \BibitemShut {NoStop}%
\bibitem [{\citenamefont {Cirelli}\ \emph {et~al.}(2011)\citenamefont
  {Cirelli}, \citenamefont {Corcella}, \citenamefont {Hektor}, \citenamefont
  {Hutsi}, \citenamefont {Kadastik}, \citenamefont {Panci}, \citenamefont
  {Raidal}, \citenamefont {Sala},\ and\ \citenamefont
  {Strumia}}]{Cirelli:2010xx}%
  \BibitemOpen
  \bibfield  {author} {\bibinfo {author} {\bibfnamefont {M.}~\bibnamefont
  {Cirelli}}, \bibinfo {author} {\bibfnamefont {G.}~\bibnamefont {Corcella}},
  \bibinfo {author} {\bibfnamefont {A.}~\bibnamefont {Hektor}}, \bibinfo
  {author} {\bibfnamefont {G.}~\bibnamefont {Hutsi}}, \bibinfo {author}
  {\bibfnamefont {M.}~\bibnamefont {Kadastik}}, \bibinfo {author}
  {\bibfnamefont {P.}~\bibnamefont {Panci}}, \bibinfo {author} {\bibfnamefont
  {M.}~\bibnamefont {Raidal}}, \bibinfo {author} {\bibfnamefont
  {F.}~\bibnamefont {Sala}}, \ and\ \bibinfo {author} {\bibfnamefont
  {A.}~\bibnamefont {Strumia}},\ }\href {\doibase
  10.1088/1475-7516/2012/10/E01} {\bibfield  {journal} {\bibinfo  {journal}
  {JCAP}\ }\textbf {\bibinfo {volume} {03}},\ \bibinfo {pages} {051} (\bibinfo
  {year} {2011})},\ \bibinfo {note} {[Erratum: JCAP 10, E01 (2012)]},\ \Eprint
  {http://arxiv.org/abs/1012.4515} {arXiv:1012.4515 [hep-ph]} \BibitemShut
  {NoStop}%
\bibitem [{\citenamefont {Stref}\ \emph {et~al.}(2019)\citenamefont {Stref},
  \citenamefont {Lacroix},\ and\ \citenamefont {Lavalle}}]{Stref:2019wjv}%
  \BibitemOpen
  \bibfield  {author} {\bibinfo {author} {\bibfnamefont {M.}~\bibnamefont
  {Stref}}, \bibinfo {author} {\bibfnamefont {T.}~\bibnamefont {Lacroix}}, \
  and\ \bibinfo {author} {\bibfnamefont {J.}~\bibnamefont {Lavalle}},\ }\href
  {\doibase 10.3390/galaxies7020065} {\bibfield  {journal} {\bibinfo  {journal}
  {Galaxies}\ }\textbf {\bibinfo {volume} {7}},\ \bibinfo {pages} {65}
  (\bibinfo {year} {2019})},\ \Eprint {http://arxiv.org/abs/1905.02008}
  {arXiv:1905.02008 [astro-ph.CO]} \BibitemShut {NoStop}%
\bibitem [{\citenamefont {{Yoshida}}(1990)}]{yoshida1990}%
  \BibitemOpen
  \bibfield  {author} {\bibinfo {author} {\bibfnamefont {H.}~\bibnamefont
  {{Yoshida}}},\ }\href {\doibase 10.1016/0375-9601(90)90092-3} {\bibfield
  {journal} {\bibinfo  {journal} {Physics Letters A}\ }\textbf {\bibinfo
  {volume} {150}},\ \bibinfo {pages} {262} (\bibinfo {year}
  {1990})}\BibitemShut {NoStop}%
\end{thebibliography}%

\end{document}